\documentclass[12pt,a4paper]{article}

\usepackage[british]{babel}
\usepackage{microtype}

\usepackage[a4paper,top=2cm,bottom=2cm,left=1.2925cm,right=1.2925cm,marginparwidth=1.75cm]{geometry}

\usepackage{amsmath}
\usepackage{graphicx}
\usepackage[colorlinks=true, allcolors=blue]{hyperref}
\usepackage{hyperref}
\usepackage[title]{appendix}
\usepackage{mathrsfs}
\usepackage{amsfonts}
\usepackage{booktabs} 

\usepackage[font=small,labelfont=bf]{caption} 

\DeclareCaptionFont{cm}{\fontfamily{cmr}\selectfont}
\usepackage{threeparttable} 
\usepackage{algorithm}
\usepackage{algorithmicx}
\usepackage{algpseudocode}
\usepackage{listings}
\usepackage{enumitem}
\usepackage{chngcntr}
\usepackage{booktabs}
\usepackage{lipsum}
\usepackage{authblk}
\usepackage[T1]{fontenc}    
\usepackage{csquotes}       
\usepackage{diagbox}

\usepackage{cleveref}

\crefname{equation}{Eq.}{Eqs.}
\Crefname{equation}{Equation}{Equation}

\crefname{figure}{Fig.}{Figs.}
\Crefname{figure}{Figure}{Figures}

\crefname{section}{Sec.}{Secs.}
\Crefname{section}{Section}{Sections}

\usepackage{siunitx} 
\DeclareSIUnit\bar{bar}

\usepackage[version=4]{mhchem}

\usepackage{setspace}
\usepackage{titlesec}
\titleformat{\section} 
  {\normalfont\Large\bfseries}{\thesection.}{1em}{}
  
\usepackage{xcolor}

\usepackage{float}   
\usepackage{caption} 


\usepackage{xspace}
\newcommand{\ie}{\textit{i.e.,}\xspace}
\newcommand{\eg}{\textit{e.g.,}\xspace}
\newcommand{\chemzip}{\texttt{ChemZIP}\xspace}

\usepackage{nomencl}
\makenomenclature

\renewcommand{\nomgroup}[1]{%
\ifthenelse{\equal{#1}{A}}{\item[\textbf{Roman Symbols}]}{%
\ifthenelse{\equal{#1}{G}}{\item[\textbf{Greek Symbols}]}{%
\ifthenelse{\equal{#1}{Z}}{\item[\textbf{Acronyms / Abbreviations}]}{%
\ifthenelse{\equal{#1}{R}}{\item[\textbf{Superscripts}]}{%
\ifthenelse{\equal{#1}{S}}{\item[\textbf{Subscripts}]}{%
\ifthenelse{\equal{#1}{H}}{\item[\textbf{Non-Dimensional Groups}]}
{}
}
}
}
}
}
}

\newcommand{\nomunit}[1]{%
 \renewcommand{\nomentryend}{\hspace*{\fill}#1}}

\newcommand{\bvec}[1]{\vec{\mathbf{#1}}}
\usepackage{bm}

\newcommand{\etal}{\textit{et al.}\xspace}
\newcommand{\kth}{$k^{\mathrm{th}}$\xspace}

\newcommand{\ecam}{\texttt{ChemZIP}\xspace}

\newcommand{\fluent}{\textsc{Fluent}\xspace}
\newcommand{\tblock}{\textsc{Tblock}\xspace}
\newcommand{\tensorflow}{\textsc{TensorFlow}\xspace}

\usepackage{subfig}
\usepackage{svg}
\usepackage{ifthen}
\usepackage{xspace}

\newcommand{\subfigbil}[1]{\textbf{(\textit{#1})}}

\renewcommand{\th}{$^{\mathrm{th}}$}

\usepackage{booktabs} 
\usepackage{multirow}
\usepackage{multicol}
\usepackage{array}
\usepackage{threeparttable}
\newcolumntype{C}[1]{>{\centering\arraybackslash}m{#1}}

\title{\chemzip: Accelerated Modeling of Complex Aerothermochemical Interactions in Novel Turbomachines for Sustainable High-Temperature Chemical Processes}

\author[1,*]{Dylan Rubini}
\author[1]{Budimir Rosic}
\affil[1]{\small Oxford Thermofluids Institute, Department of Engineering Science, University of Oxford, OX1 3PJ, UK}
\affil[*]{Corresponding author: \texttt{dylan.rubini@eng.ox.ac.uk}}

\date{}  

\begin{document}
\maketitle

\begin{abstract}

\noindent \textit{This paper introduces a new platform to accelerate the modeling of complex aerothermochemical interactions in new turbomachines, called turbo-reactors, to decarbonise chemical processes. While previous work has aerothermally demonstrated the potential to decarbonize the heat input to the reaction, optimizing the reaction efficiency has been a challenge. This is because measuring reaction performance with aerochemical simulations is computationally prohibitive due to the uniquely complex aerodynamics and chemistry within turbomachines. To address this, we introduce a new multifidelity machine-learning-assisted methodology, called \chemzip, to mitigate this bottleneck. Although data-driven methodologies exist for combustion, modeling reactive flows along the bladed path of a turbomachine poses new challenges. This has led to a novel training data generation process, which allows rich dynamic responses of the chemical system to be embedded into the training dataset at a fraction of the cost of reacting flow simulations. The resulting high-dimensional composition vector is compressed into a low-dimensional basis using an autoencoder-like neural network, inspired by but more universal than traditional flamelet-generated manifolds. Verification against 10,000 unseen one-dimensional test conditions shows an $R^2$ score exceeding 95\% across all quantities of interest. Following this, \chemzip is coupled into a fully-fledged viscous computational fluid dynamics solver. For a set of process-relevant three-dimensional configurations entirely different from the training data, the predictive accuracy of the thermochemical state remains within 10\% of an industry-standard solver (\fluent) while convergence is achieved 50 times faster, even for a small mechanism. Therefore, numerical computations are sufficiently fast that aerothermo\textit{chemical} optimization is now feasible for the first time in the design cycle of the turbo-reactor.} 

\end{abstract}


\vspace*{2em}
\noindent \textbf{Highlights}\\
\begin{itemize}

  \item Complex aerothermochemical interactions exist in a new class of turbomachines.
  \item A machine-learning-assisted platform for accelerating reactive flows is introduced.
  \item ChemZIP is 50 times faster than state-of-the-art acceleration methods in Fluent.
  \item The accuracy is within 10\% of Fluent for a long 3D heated duct domain.
  \item ChemZIP will enable aerochemically-guided design optimization for the first time.

\end{itemize}



\nomenclature[A-$t$]{$t$}{Time \nomunit{\si{\second}}}
\nomenclature[A-$Y_k$]{$Y_k$}{Mass fraction of the \kth species \nomunit{\%}}
\nomenclature[A-$Y$]{$\boldsymbol{Y}$}{Mass fraction  vector $\in \mathbb{R}^{K}$ \nomunit{\%}}

\nomenclature[A-$Q$]{$\dot{Q}$}{Imposed heat flux \nomunit{\si{\kilo\watt\per\meter\squared}}}

\nomenclature[A-$D_k$]{$\bar{D}_k$}{Molecular diffusivity of \kth species \nomunit{\si{\metre\squared\per\second}}}
\nomenclature[A-$D_{\mathrm{eff}}$]{$\boldsymbol{D}_{\mathrm{eff}}$}{Total diffusivities 
 matrix $\in \mathbb{R}^{K \times K}$  \nomunit{\si{\metre\squared\per\second}}}

\nomenclature[A-$D_{Z}$]{$\boldsymbol{D}_{Z}$}{Meta species diffusivities matrix $\in \mathbb{R}^{M \times M}$ \nomunit{$-$}}

\nomenclature[A-$R^2$]{$R^2$}{Coefficient of determination \nomunit{$-$}}
\nomenclature[A-$T$]{$T$}{Static temperature \nomunit{\si{\kelvin}}}
\nomenclature[A-$p$]{$p$}{Static pressure \nomunit{\si{\pascal}}}
\nomenclature[A-$c_p$]{$c_p$}{Isobaric specific heat capacity \nomunit{\si{\joule\per\kilogram\per\kelvin}}}

\nomenclature[A-$u$]{$\bvec{u}$}{Velocity vector \nomunit{\si{\meter\per\second}}}
\nomenclature[A-$R$]{$R$}{Specific gas constant \nomunit{\si{\joule\per\kilogram\per\kelvin}}}

\nomenclature[A-$K$]{$K$}{Number of species \nomunit{$-$}}
\nomenclature[A-$J$]{$\mathbf{J}$}{Species Jacobian matrix $\in \mathbb{R}^{K \times K}$ \nomunit{\si{\per\second}}}

\nomenclature[A-$J$]{$\mathbf{J}_{\phi}$}{Encoder Jacobian matrix $\in \mathbb{R}^{M \times K}$ \nomunit{$-$}}
\nomenclature[A-$J$]{$\mathbf{J}_{\psi}$}{Decoder Jacobian matrix $\in \mathbb{R}^{K \times M}$ \nomunit{$-$}}

\nomenclature[A-$Z_k$]{$Z_k$}{Mass fraction for the $m^{\text{th}}$ latent meta species \nomunit{$-$}}
\nomenclature[A-$Z$]{$\boldsymbol{Z}$}{Meta species mass fractions vector $\in \mathbb{R}^{M}$  \nomunit{$-$}}

\nomenclature[A-$M$]{$M$}{Number of latent meta species \nomunit{$-$}}

\nomenclature[A-$N$]{$N$}{Number data points \nomunit{$-$}}

\nomenclature[A-$X$]{$\mathbf{X}$}{Input space training data matrix $\in \mathbb{R}^{N \times K+2}$ \nomunit{$-$}}
\nomenclature[A-$W$]{$\mathbf{W}$}{Output dynamic response training data matrix $\in \mathbb{R}^{N \times K+4}$ \nomunit{$-$}}

\nomenclature[A-$F_{NN}$]{$\mathscr{F}_{\mathrm{NN}}$}{Neural network function approximator \nomunit{$-$}}

\nomenclature[A-$F_{NN,Z}$]{$\mathscr{F}_{\mathrm{NN,Z}}$}{Latent dynamic response approximator \nomunit{$-$}}


\nomenclature[G-$\alpha_T$]{$\alpha_T$}{Molecular thermal diffusivity \nomunit{\si{\meter\squared\per\second}}}

\nomenclature[G-$\mu$]{$\mu$}{Molecular/laminar viscosity \nomunit{\si{\pascal\second}}}
\nomenclature[G-$\mu_t$]{$\mu_t$}{Turbulent viscosity \nomunit{\si{\pascal\second}}}
\nomenclature[G-$\rho$]{$\rho$}{Density \nomunit{\si{\kilogram\per\meter\cubed}}}
\nomenclature[G-$\omega_k$]{$\dot{\omega}_k$}{Net production rate of \kth species \nomunit{\si{\per\second}}}

\nomenclature[G-$\omega$]{$\boldsymbol{\dot{\omega}}$}{Species net production rate vector $\in \mathbb{R}^{K}$  \nomunit{\si{\per\second}}}

\nomenclature[G-$\omega_{Z_m}$]{$\dot{\omega}_{Z_m}$}{Net production rate of the $m^{\text{th}}$ meta species \nomunit{$-$}}

\nomenclature[G-$\omega_{Z_m}$]{$\boldsymbol{\dot{\omega}}_{Z}$}{Net production rate vector for the latent meta species $\in \mathbb{R}^{K}$ \nomunit{$-$}}

\nomenclature[G-$\tau_{f}$]{$\tau_{f}$}{Characteristic fluid time-scale \nomunit{\si{\second}}}
\nomenclature[G-$\tau_{c_k}$]{$\tau_{c_k}$}{Characteristic chemical time-scale \nomunit{\si{\second}}}

\nomenclature[G-$\omega_T$]{$\dot{\omega}_T$}{Reaction heat release/absorption rate \nomunit{\si{\watt\per\cubic\meter}}}


\nomenclature[Z-NMdAE]{NMdAE}{Normalised median absolute error}
\nomenclature[Z-NRMSE]{NRMSE}{Normalised root mean square error}
\nomenclature[Z-MAPE]{MAPE}{Mean absolute percentage error}
\nomenclature[Z-LUT]{LUT}{lookup table}
\nomenclature[Z-FGM]{FGM}{Flamelet generated manifold}
\nomenclature[Z-ISAT]{ISAT}{\textit{In situ} adaptive tabulation}
\nomenclature[Z-DCZ]{DCZ}{Dynamic cell zoning}
\nomenclature[Z-DI]{DI}{Direct integration (of the chemical kinetics)}
\nomenclature[Z-DACM]{DACM}{Dynamic adaptive chemistry method}

\nomenclature[Z-CFD]{CFD}{Computational fluid dynamics}
\nomenclature[Z-RANS]{RANS}{Reynolds-averaged Navier-Stokes}

\nomenclature[Z-PFR]{PFR}{Plug-flow reactor}
\nomenclature[Z-FOM]{FOM}{Full-order model}
\nomenclature[Z-ML]{ML}{Machine learning}

\nomenclature[Z-NN]{NN}{Neural network}
\nomenclature[Z-ANN]{ANN}{Artificial neural network}
\nomenclature[Z-PCA]{PCA}{Principal component analysis}
\nomenclature[Z-AE]{AE}{Autoencoder}
\nomenclature[Z-MSE]{MSE}{Mean square error}
\nomenclature[Z-TF]{TF}{\tensorflow}
\nomenclature[Z-TCI]{TCI}{Turbulence-chemistry interactions}

\nomenclature[R-$\square^{\prime}$]{$\square^{\prime}$}{Fluctuating component}
\nomenclature[R-$\widehat{\square}$]{$\widehat{\square}$}{Reconstructed/predicted quantity}

\nomenclature[S-$\square_{k}$]{$\square_{k}$}{Species index}
\nomenclature[S-$\square_{j}$]{$\square_{j}$}{Reaction index}

\nomenclature[S-$\square_{m}$]{$\square_{m}$}{Latent meta species index}

\nomenclature[R-$\overline{\square}$]{$\overline{\square}$}{(Time/Reynolds) averaging}
\nomenclature[R-$\bar{\square}$]{$\bar{\square}$}{Mixture averaging}


\nomenclature[H-$Sc_t$]{$Sc_t$}{Turbulent Schmidt number}
\nomenclature[H-$Le_k$]{$Le_k$}{Lewis number for \kth species}

\printnomenclature[2.0cm]

\section{Introduction}

\subsection{Motivation and Background}

\begin{figure*}[htb]
    \centering
    \includegraphics[width=160mm]{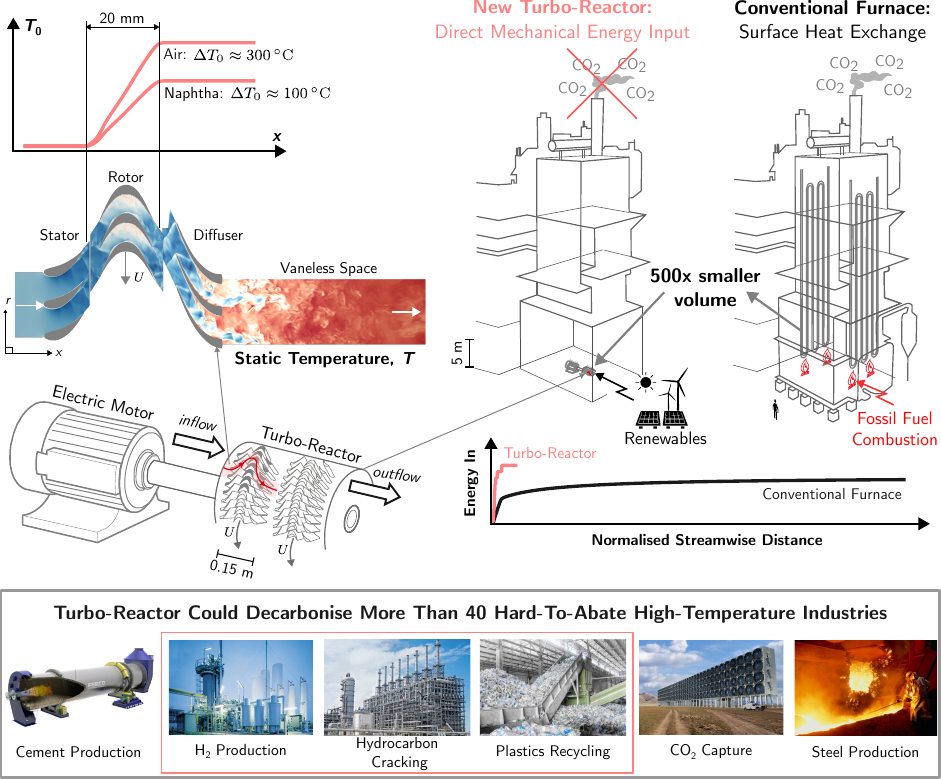} 
    \caption{Electrified turbo-reactor could decarbonise many hard-to-abate industrial processes.}
    \label{fig:RDRVsHeX}
\end{figure*}

The turbomachinery-based turbo-reactor concept shown in \cref{fig:RDRVsHeX} is capable of decarbonizing the fired process heat within 40 hard-to-abate, high-temperature ($\SI{400}{\celsius} \leq T \leq \SI{1700}{\celsius}$) industrial processes ranging from steel and cement to chemicals and plastic recycling. This singular turbomachinery concept has the potential to eliminate a large fraction of greenhouse gas emissions due to industrial energy consumption, which was around 25\% of global emissions in 2020 \cite{owid-ghg-emissions-by-sector}. 

Previous work demonstrated the feasibility of transferring a process-specific amount of energy to a range of working fluids over a distance 10 to 20 times less than an axial compressor without system instability \cite{DylanASME21, DylanGPPS21, NKDylanASME23, DylanGPPS23,DylanASME22}. This was shown purely from an \textit{aerothermal} perspective using nonreacting low/high fidelity numerical simulations. However, in many applications of the turbo-reactor, the working fluid undergoes a thermochemical decomposition reaction (\eg methane reforming, pyrolysis, ammonia cracking, etc.) to produce valuable products at the outlet while trying to minimize unwanted secondary products. Therefore, aerochemical interactions along the bladed path are fundamental to the process for the first time in turbomachinery. 


As explained in Rubini, Rosic \& Xu \cite{Rubini2024}, there is now a new turbomachinery design landscape to go beyond traditional aerothermal design optimization and instead aerothermo\textit{chemically} optimize the stage design to produce a reaction-efficient temperature distribution. However, although this is vital for the success of the turbo-reactor concept, aerothermochemical design optimization has been left virtually unexplored because of the prohibitive computational burden of reacting flow simulations using realistic, detailed, and comprehensive chemical kinetic models. Even for a simple kinetic model with 65 species simulated on a coarse 1-million-node mesh, a reacting flow solution can take $\sim$3 days to converge on a typical workstation despite employing state-of-the-art acceleration strategies for reacting flows. Now imagine querying an objective function at least 100 to 1000 times in a multi-query optimization/exploration outer loop. This would take years, which is by definition intractable. 

Moreover, in practice, the goal is to accommodate more comprehensive and thus realistic kinetic models \cite{LU2009192} to improve the accuracy of reaction modeling under a wide range of conditions. For propane pyrolysis for the production of olefins \cite{FAKHROLESLAM2019553,van2010next}, a detailed mechanism can contain more than 900 species and almost 10,000 reactions \cite{D2FD00032F} (which is, in fact, 4 to 5 times larger than a typical detailed kinetic model of natural gas combustion \cite{FRASSOLDATI200397}). This issue of dimensionality is exacerbated by the use of high carbon number feedstocks in the chemical industry, such as naphtha (which contains molecules with up to 12 carbon atoms). 

Even with current state-of-the-art acceleration strategies available for reacting flows in chemical engineering, such as \textit{ in situ} adaptive tabulation (ISAT) \cite{doi:10.1080/713665229}, incorporating more than 10 to 20 species is not feasible at the industrial design level \cite{westbrook2008role}. On the other hand, in numerical combustion modeling, the flamelet-generated manifold (FGM) approach \cite{doi:10.1080/00102200008935814} has proven to be a highly efficient compression mechanism for flame calculations. However, in its current form, it is only applicable to combustion and is therefore not universal.

\begin{figure}[htb]
    \centering
    \includegraphics[width=88mm]{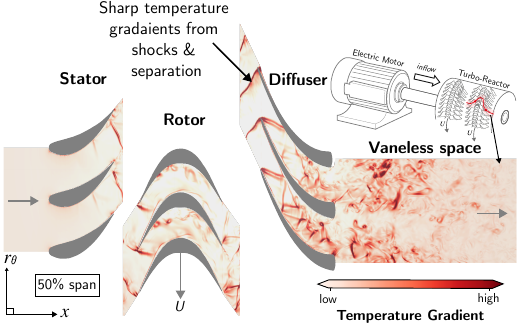} 
    \caption{High-fidelity numerical simulation showing the instantaneous temperature gradients present in the flow.}
    \label{fig:UniqueFlowPhysics2}
\end{figure}

\begin{figure}[htb]
    \centering
    \includegraphics[width=88mm]{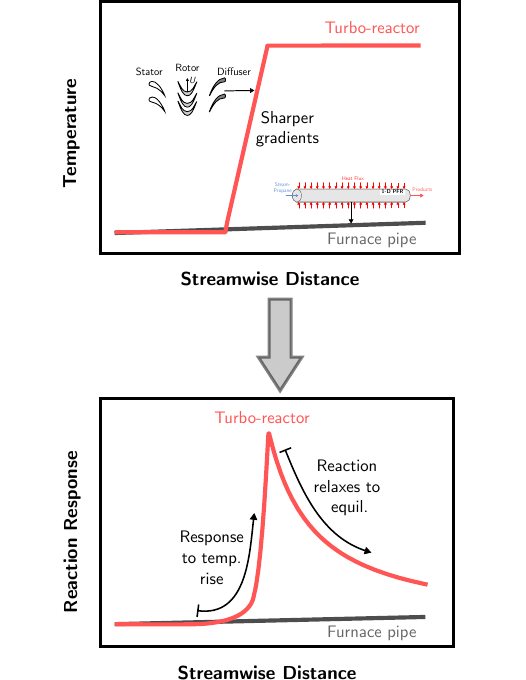} 
    \caption{Schematic illustration of the reaction response to representative temperature gradients in the turbo-reactor compared to those in a tubular pipe of a conventional furnace.}
    \label{fig:reaction_regimes2}
\end{figure}

Consequently, optimization of the reaction performance in the turbo-reactor has so far been restricted to simple, idealized, homogeneous 1-D chemical kinetic calculations. Although this can give realistic estimates for simple flows such as tubular chemical furnaces \cite{westbrook2008role}, the same cannot be said for complex aerothermal phenomena observed in the turbo-reactor. As a result of the extreme power density, \cref{fig:UniqueFlowPhysics2} shows strong temperature gradients in the flow much higher in intensity than traditional furnace-based chemical devices. These are due to flow features such as shock systems and flow separations. As a knock-on effect, the reaction operates in a highly nonequilibrium state (see \cref{fig:reaction_regimes2}), adding a further layer of complexity. Therefore, to date, more accurate predictions of basic performance metrics have only been available after costly and time-consuming physical experiments---and these are severely limited by low spatial resolution, so cannot guide design optimization.

The aforementioned limitations of the current design system can lead to disastrously sub-optimal reaction efficiency measured for the manufactured machine (relative to the design intent) and result in expensive iterative re-design and re-manufacturing. Given the accelerating climate emergency and the urgency for the net-zero transition, this laborious process is not sustainable.

\subsection{Contributions in this Paper}

To overcome the limitations of traditional physics-based computational acceleration strategies (\eg ISAT) for reactive flows, this paper introduces a new multifidelity efficiently-coupled machine-learning-enhanced (ML) aerochemical modeling methodology called \chemzip. This approach is more scalable to detailed kinetics than ISAT and more universal than FGM (applicable to both processes in chemical engineering and combustion), and the computational time for convergence is orders of magnitude faster than that of traditional approaches. Despite some loss of absolute accuracy, this is acceptable for design optimization if trends are well predicted. 

Although ML approaches have been proposed for numerical combustion modeling \cite{BLASCO199838,ZHANG2022112319,GOSWAMI2024116674,SUTHERLAND20091563,MALIK20212635,MAO2023108842}, \chemzip introduces new contributions to the field due to the novelty of the application. To capture the unique level of complexity within the flow (see \cref{fig:UniqueFlowPhysics2,fig:reaction_regimes2}), this methodology offers a new strategy to embed the dynamic response of complex flow physics and detailed chemistry into the training database within only a few hours. 


The outline of the paper is as follows. \Cref{sec:turbo_concept}, presents the basic working principles of the turbo-reactor concept for decarbonizing chemical processes. \Cref{sec:model_aerochem_interaction} presents the challenges of modeling reactive flows in the turbo-reactor before highlighting the limitations of both traditional physics- and data-driven acceleration techniques. \Cref{sec:method} explains the new \chemzip methodology. Following this, \Cref{sec:results} presents a detailed verification and performance benchmarking study comparing \chemzip against an industry-standard computational fluid dynamics (CFD) solver, \fluent, within a multidimensional viscous flow environment. Finally, a first-of-its-kind proof-of-concept numerical simulation of aerochemistry along the bladed path of a turbomachine (\ie the turbo-reactor) is showcased.

Although the \chemzip platform is introduced in the context of conducting endothermic gas-phase chemistry within a turbomachine, it can also be applied to a wide variety of applications where high-dimensional chemistry is the computational bottleneck. This would require additional physics to be added such as photochemistry for atmospheric chemistry \cite{10.1093/mnras/stad1763}, charge transfer reactions for electrochemical devices \cite{kee2017chemically}, heterogeneous catalytic surface reactions for catalytic reactors \cite{D3RE00212H}, and electronic impact energy transfer reactions for plasma applications
\cite{Peerenboom_2015}.

\section{The Ultra-High Power Density Turbo-Reactor Concept} \label{sec:turbo_concept}
Direct electrification is likely to be the most cost-effective solution for decarbonizing hard-to-abate processes such as hydrocarbon \& ammonia cracking and sustainable feed/fuel production. Burning hydrogen is too expensive \cite{SHAFIEE2024} and high-temperature, high-capacity electric resistance furnaces have severe limitations such as oxidation and an irreducibly large footprint. Therefore, the turbo-reactor concept has been introduced to replace high-temperature industrial furnaces with an electric-motor-driven turbomachine powered by renewable electricity (see \cref{fig:RDRVsHeX,fig:UniqueFlowPhysics2}). Direct mechanical energy transfer in the turbo-reactor---equivalent to volumetric heating---enables a 500 times increase in power density relative to surface heat exchange in conventional furnaces. By converting almost all of the mechanical energy imparted into static temperature rather than pressure of the fluid, an enthalpy input rate 10 to 20 times higher than conventional energy-imparting turbomachines (\ie compressors) is possible. Volumetric heating combined with streamwise temperature profile controllability, extreme turbulent mixing, lateral temperature uniformity, and a short reaction time scale enables substantial improvements in reaction performance, contributing to a more cost-effective process. For more detailed on the working principles of the turbo-reactor concept, the reader is refereed to Refs. \cite{Rubini2024, DylanASME21,
DylanGPPS21, NKDylanASME23}

\section{Modeling Aerochemical Interactions} \label{sec:model_aerochem_interaction}

\subsection{Challenges}

Modeling nonequilibrium finite rate chemistry \cite{kee2017chemically,Oran_Boris_2000} is one of the biggest challenges in computational science for two reasons. First, the problem is high-dimensional with an additional $K$ nonlinearly coupled species conservation equations. In many chemical processes such as olefin production, $K$ can reach 900 for accurate kinetics \cite{D2FD00032F}. Second, chemical kinetics are highly nonlinear, multiscale, and numerically stiff with characteristic chemical timescales ranging from seconds to picoseconds \cite{FRASSOLDATI200397}. Therefore, stiff implicit time integrators are necessary to correctly scale the effective time step for each species \cite{kee2017chemically}. The implicit methods require inversion of the Jacobian matrix $\mathbf{J} \in \mathbb{R}^{K \times K}$. Since this operation scales cubically with the size of the matrix $\mathcal{O}(K^3)$, modeling detailed chemistry is intractable \cite{LU2009192}. These challenges are exacerbated by modeling the complex flow physics in the turbo-reactor. This is because intense gradients in the flow mean (see \cref{fig:UniqueFlowPhysics2,fig:reaction_regimes2}) that more complex kinetic models and higher resolution computational grids are necessary. 

\subsection{Limitations of Existing Accelerations Approaches}

In most industrial applications, direct integration (DI) of the reaction network is replaced by physics-based computational acceleration strategies introduced below. However, these approaches are inadequate for many applications, either due to insufficient speedup or inadequate universality.

\subsubsection{Physics-Driven} \label{sec:physics_based}

For complex feedstocks used in the chemical industry, such as (bio/synthetic) naphtha or fuels for combustion, skeletal mechanisms produced by mechanism reduction \cite{5564c59f373b4baf9a59314ec833bd95,PEPIOTDESJARDINS200867} can have 50 to 200 species, which are still prohibitively costly for 3D viscous reacting flow simulations for an industrial setting. A somewhat greater level of computational acceleration can be achieved by building reduced kinetic models ``on the fly''  locally within each cell of the mesh using the dynamic adaptive chemistry method (DACM) from Liang, Stevens \& Farrell \cite{LIANG2009527}; however the speedup is still not enough for many problems.

To alleviate the overhead of evaluating the chemical source terms $\dot{\omega}_k$ and inverting the Jacobian matrix, Pope \cite{doi:10.1080/713665229} introduced the ISAT method to dynamically grow a look-up table (LUT) in real-time in each mesh cell with a specified error tolerance.  ISAT only works well for small mechanisms of less than 50 species because storage and retrieval costs scale quadratically $\mathcal{O}(K^2)$ with the number of species \cite{doi:10.1080/13647830500307501} and it does not alleviate the large number of transport equations. 

This challenge can be addressed using FGM, which projects the high-dimensional thermochemical state vector onto a two-dimensional basis formed by the mixture fraction and the reaction progress variables \cite{doi:10.1080/00102200008935814,6c7475beb5f541a08899d1357b36e77f}. However, this method relies on the notion of a flamelet and requires process-specific expert knowledge to define the reduced-dimensionality state space.

\subsection{Data-driven Acceleration Approaches}
\subsubsection{Why Machine Learning?}

\begin{figure}[htb]
    \centering
    \includegraphics[width=88mm]{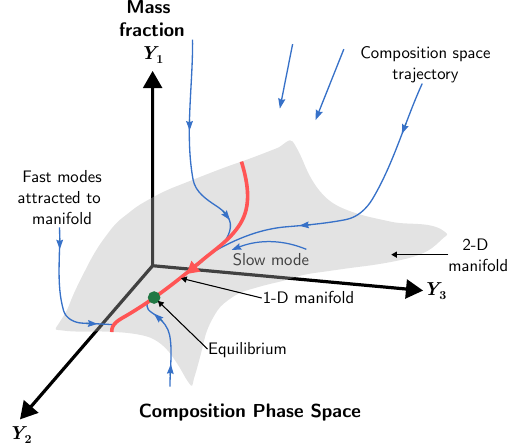} 
    \caption{Schematic illustration of a low-dimensional manifold in composition phase space.}
    \label{fig:manifold}
\end{figure}

By using a \textit{machine-learning-assisted} methodology, it is possible to learn a low-dimensional thermochemical latent space (similar to FGM) automatically and for an arbitrary reaction process (not just combustion), while simultaneously learning an ultrafast approximation for the chemical source term mappings. The rich and flexible nonlinear representational capacity of neural networks can overcome the \textit{curse of dimensionality} \cite{Goodfellow-et-al-2016} of tabulation-based methods and the lack of functional expressiveness of reduced skeletal mechanisms, in addition to providing a level of universality that is not possible with FGM. Fundamentally, compression of high-dimensional chemical systems is justified by the fact that trajectories in phase space are naturally attracted to low-dimensional manifolds, as shown in \cref{fig:manifold}, suggesting reducible redundancy in the governing physics. 

\begin{figure*}[htb]
    \centering
    \includegraphics[width=183mm]{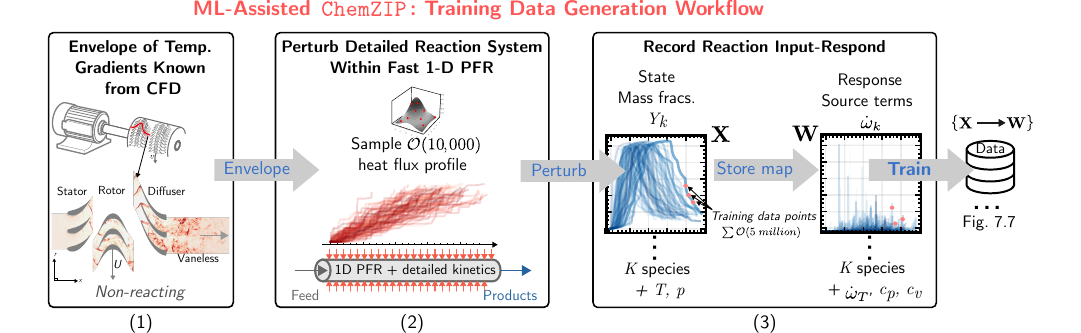} 
    \caption{Schematic of the training data generation workflow for the \chemzip methodology.}
    \label{fig:NeuralNetWorkFlow_training}
\end{figure*}

\subsubsection{Existing Implementations}

Since the 1990s artificial neural networks (ANN) have been used to approximate the costly functional mapping between the net species production rate vector $\boldsymbol{\dot{\omega}} =  [\dot{\omega}_1, \ldots, \dot{\omega}_K]^{\top}$ and the static temperature $T$, static pressure $p$, and mass fraction vector $\boldsymbol{Y} = [Y_1, \ldots, Y_K]^{\top}$:
\begin{equation} \label{eq:nn_map}
    \boldsymbol{\dot{\omega}} \approx \mathscr{F}_{\mathrm{NN}}(T, p, \boldsymbol{Y}),
\end{equation}
where $\mathscr{F}_{\mathrm{NN}}$ represents the neural network (NN) mapping operator. Today, neural networks are used extensively for chemical systems, as illustrated in Refs. \cite{CHRISTO199643,BLASCO199838,annvsisat,CHEN2000115,SEN201062,peng2017efficient,MAO2023108842,D3RE00212H,ZHANG2022112319}. However, in their most basic form, ANNs do not eliminate numerical stiffness or address the large number of transport equations for detailed kinetic models. Neural operators \cite{GOSWAMI2024116674} have provided a solution for the first problem and principal component analysis (PCA) or autoencoder NNs to the latter \cite{SUTHERLAND20091563,MIRGOLBABAEI2014118,ARMSTRONG2024113119,MALIK20212635,ZDYBAL2023100859,VIJAYARANGAN2024100325,KUMAR2024131212,Zhang2022,refId0}. Additionally, neural ordinary differential equation (ODE) solvers have also been widely used as an efficient way to combine the ODE integrator with the NN nonlinear regression approximation of the source term \cite{OWOYELE2022100118}.

\section{ChemZIP: An Aerochemical Modeling Methodology} \label{sec:method}

In this section, the implementation details of the new \chemzip methodology will be described, including the generation of training data, the neural network architecture, and the treatment of kinetic-fluid coupling. The new \ecam methodology uses a NN architecture to compress the high-dimensional thermochemical space $\mathcal{O}(200 - 1000)$ onto a substantially lower-dimensional manifold $\mathcal{O}(3)$, embedded in the original state space. Here, the dynamics of the system can be tracked efficiently. In latent space, the reduced set of \textit{metaspecies} mass fractions are mapped to their corresponding source terms, which correspond to a projection of the physical dynamic response into latent space. The approximate chemical system is then coupled with an arbitrary-fidelity flow solver without sacrificing the accuracy of the preexisting aerothermal transport modeling.

So far in this work, simpler implementation approaches have been adopted since the objective of this paper primarily is to prove the viability of the methodology.

\subsection{Training Data Generation}

\begin{figure*}[htb]
    \centering
    \includegraphics[width=183mm]{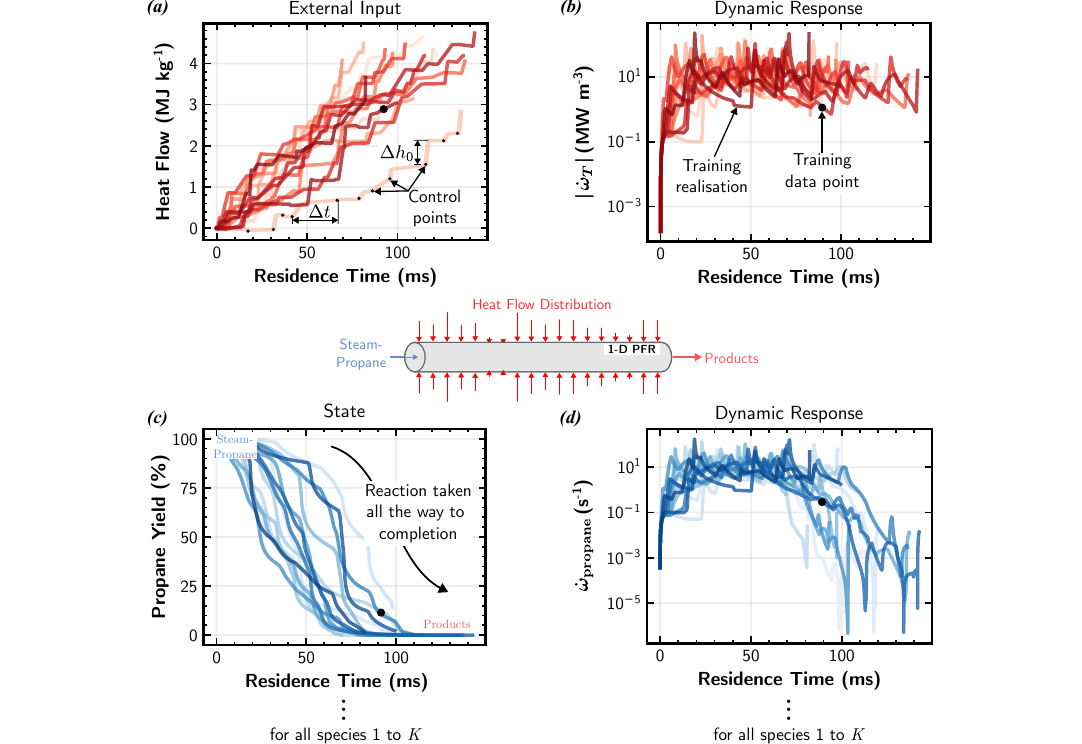}
    \caption{A selection of 10 (out of 10,000) example training data 1-D PFR reactor simulations showing \subfigbil{a} the heat flow \subfigbil{b} the reaction heat absorption rate on a log scale \subfigbil{c} the dry propane yield and \subfigbil{d} the propane net production rate on a log scale.}    
    \label{fig:training_data_profiles}
\end{figure*}



The quality of any machine learning method usually boils down to the quality and size of the suite training data. In computational science, training data generation is often the bottleneck of an ML pipeline. The \ecam methodology addresses this challenge by simplifying the chemistry model down to a perfectly-mixed, constant-area 1-D plug-flow reactor (PFR) while maintaining an accurate and detailed kinetic model with $\mathcal{O}(200 - 1000)$ species. By reducing the cost of a simulation, tens of thousands of different conditions for the detailed kinetic model can be predicted in just a few hours (on an 18-core CPU), thus eliminating the training data generation bottleneck.

Neural networks are ultimately just universal input--output function approximators. Therefore, even in a simplified 1D environment, if we generate training data in such a way that all possible realizable combinations of the \textit{inputs} $Y_k$, $T$ \& $p$ are mapped to the dynamic response (\textit{outputs}) of the reaction network in terms of the net species production rates $\dot{\omega}_k$, as well as $\dot{\omega}_T$, $c_p$, $R$, \& the molecular viscosity $\bar{\mu}$, then in theory, with a sufficiently large NN function approximator, any thermochemical input state could be accurately mapped to the output reaction response. However, to make this problem tractable, the phase-space exploration is constrained to turbo-reactor-relevant conditions. In order to capture the wide range of complex dynamic response regimes (see \cref{fig:UniqueFlowPhysics2,fig:reaction_regimes2}) within the turbo-reactor, first, a nonreacting precursor CFD simulation is conducted to determine the envelope of realistic temperature gradients seen by a fluid parcel traveling along the bladed flow path (\eg across a shock). This forms the first component of the overall training data generation workflow shown in \cref{fig:NeuralNetWorkFlow_training}. 

The next step is to randomly sample several thousand heat flux profiles within the envelope of temperature gradients set by the precursor nonreacting CFD (see \cref{fig:NeuralNetWorkFlow_training}). By perturbing the reaction through sharp gradients in heat input to the reaction, the dynamic response of the reaction system ($\dot{\omega}_k$, $\dot{\omega}_T$, $c_p$, etc.) can be explored and recorded under a wide range of thermochemical conditions (see \cref{fig:NeuralNetWorkFlow_training}). 

By repeatedly perturbing the reaction at different rates and at different times, the sensitivities (transfer functions) of the reaction system are generated; this should allow the downstream learning algorithm to construct a manifold that approximates the high-dimensional trajectory through the composition space (see \cref{fig:manifold}). A key enabling assumption of this approach is that the reaction is strongly temperature-driven rather than mixing-driven like in many combustion problems.

The method described above is agnostic to the kinetic model and/or initial composition used. To the best of our knowledge, this perturbation-based training data generation strategy has not yet been leveraged in the literature. 

\subsubsection{Sample Space}

The shape of each heat flux profile is set by $N_{\text{control}}$ control points (see \cref{fig:training_data_profiles}(\textit{a})), where $N_{\text{control}}$ is typically around 10 to 20. At each independent control point, the following control parameters can be varied to shape the heat flux profile: the residence time between control points ($\Delta t$), the heat input level ($\Delta h_0$), the heat input ramp rate, and the heat loss level (see \cref{fig:training_data_profiles}). The heat input can be either positive or negative to simulate both heating and cooling. In addition, the pressure level can also float. If $N_{\text{control}} = 15$, this means that the total dimensionality of the sample space would be $15 \times 5 = 75$-D.

\begin{figure*}[htbp]
    \centering
    \includegraphics[width=183mm]{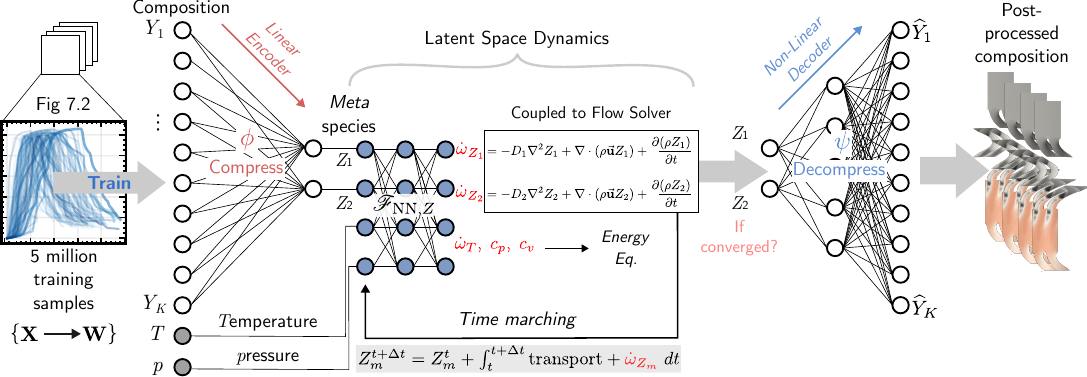} 
    \caption{ Schematic illustration showing the \chemzip chemical autoencoder NN architecture for compressing the thermochemistry and coupling with a 3-D viscous CFD flow solver.}
    \label{fig:autoencoder}
\end{figure*}

Sobol sampling \cite{SOBOL196786} is used to distribute the samples uniformly in this high-dimensional (\eg 75-D) space. On the order of 5,000 to 10,000 heat flux profiles are randomly generated. To diversify the sample space, 2,500 constant-heat-flux samples are added where only the magnitude is randomly varied on a log scale. For control parameters that are allowed to change over many orders of magnitude, such as the ``residence time between control points'', sampling can be performed on a logarithmic (rather than linear) scale to better spread the samples \cite{ZHANG2022112319}.

\subsubsection{One-Dimensional Plug Flow Reactor Model}

For each heat flux profile realization, a constant-pressure 1-D PFR simulation is run (see \cref{fig:NeuralNetWorkFlow_training,fig:training_data_profiles}) using \textsc{Cantera} \cite{cantera}. For each realization, the simulation is shutdown early if the temperature exceeds a user-defined limit of $T_{\mathrm{max.}} = \SI{1150}{\celsius}$, which is 15\% higher than the maximum allowable operating temperature of the turbo-reactor. 

To allow a larger number of training data points to be clustered around regions of more complex reaction dynamics (which are thus challenging to capture), \textsc{Cantera} is set to take adaptive time steps. The underlying time integrator naturally distributes more time steps in regions of high gradients where the truncation error rises above a threshold. Rather than storing the data for each time step, the data is subsampled every two to four steps.


\subsection{The Training Database}

\newcommand{\mymatrix}[1]{\ensuremath{\left\downarrow\vphantom{#1}\right.\overset{\xrightarrow[\hphantom{#1}]{\text{state variables}}}{#1}}}

For each heat flux profile, $\mathcal{O}(1000)$ time steps are taken and the thermochemical state and rich dynamic response of a detailed mechanism are stored in a training database. The input state for the training database can be written as $\mathbf{X} \in \mathbb{R}^{N} \times \mathbb{R}^{K+2}$, where
\begin{equation}
\mathbf{X} = 
\text{\tiny time}\mymatrix{\begin{bmatrix}
Y_1(t_1) & Y_2(t_1) & \cdots & Y_K(t_1) & T(t_1) & p(t_1) \\
Y_1(t_2) & Y_2(t_2) & \cdots & Y_K(t_2) & T(t_2) & p(t_2) \\
\vdots & \vdots & \ddots & \vdots & \vdots & \vdots \\
Y_1(t_N) & Y_2(t_N) & \cdots & Y_K(t_N) & T(t_N) & p(t_N)
\end{bmatrix}},
\label{eq:train_data_matrix}
\end{equation}
and the output response $\mathbf{W} \in \mathbb{R}^{N} \times \mathbb{R}^{K+4}$ as
\begin{equation}
\mathbf{W}
= \begin{bmatrix}
\dot{\omega}_1(t_1) & \cdots & \dot{\omega}_K(t_1) & \dot{\omega}_T (t_1) & c_p(t_1) & c_v(t_1)   \\
\dot{\omega}_1(t_2)  & \cdots & \dot{\omega}_K(t_2) & \dot{\omega}_T(t_2) & c_p(t_2) & c_v(t_2)  \\
\vdots & \ddots & \vdots & \vdots & \vdots & \vdots \\
\dot{\omega}_1(t_N) & \cdots & \dot{\omega}_K(t_N) & \dot{\omega}_T(t_N) & c_p(t_N) & c_v(t_N)  )
\end{bmatrix}.
\label{eq:train_data_response_matrix}
\end{equation}
where $\dot{\omega}_T$ is the reaction heat adsorption rate, $c_p$ is the isobaric heat capacity, $c_v$ is the isochoric heat capacity. If necessary, molecular viscosity can also be added to the output response matrix. By applying the \textit{frozen chemistry} assumption, other thermodynamic properties can be calculated from $c_p$ and $c_v$. The temporal snapshots stored for each heat flux profile realization are stacked vertically in \cref{eq:train_data_matrix,eq:train_data_response_matrix}. Therefore, the training library maps all composition-dependent variables from the thermochemical input state to their output response $\{\mathbf{X} \mapsto \mathbf{W}\}$ (see \cref{fig:NeuralNetWorkFlow_training}). This forms a large corpus of training data, where $N = \mathcal{O}(5 \text{ million})$ samples for each variable.

\subsection{Data-Driven Kinetics Compression}

Once the training database has been compiled (see \cref{eq:train_data_matrix,eq:train_data_response_matrix}), \cref{fig:autoencoder} shows how machine learning is exploited to compress the high-dimensional thermochemical space onto a low-dimensional manifold using an encoder---$\phi : \mathbb{R}^K \mapsto \mathbb{R}^M, \text{where}\\ M \ll K$---while simultaneously learning a regression mapping for the dynamic response in latent space $\boldsymbol{\dot{\omega}}_Z = \mathscr{F}_{NN,Z}(T, p, \boldsymbol{Z})$. Although not strictly correct, it is helpful to think of the latent space basis as a set of ``meta'' or artificial species. During training, a mapping $\mathbf{J}_{\phi}$ between the metaspecies mass fractions $\boldsymbol{Z} = [Z_1, Z_2, \ldots, Z_M]^{^{\top}}$, physical temperature \& pressure, and the reaction rates projected into the latent space $$\boldsymbol{\dot{\omega}}_Z = [\dot{\omega}_{Z_1}, \dot{\omega}_{Z_2}, \ldots, \dot{\omega}_{Z_M}]^{\top} = \mathbf{J}_{\phi}\boldsymbol{\dot{\omega}}$$ is learned (see \cref{fig:autoencoder}). The source terms $\boldsymbol{\dot{\omega}}_Z$---a lumped representation of the dynamic response in latent space---are coupled with a convection-diffusion transport equation in a standard CFD flow solver. Time integration can then be performed in reduced latent space, before the full multi-species spatiotemporal distribution of mass fractions in physical space $\widehat{Y}_k$ is recovered/reconstructed during post-processing.

The following subsections explain each of the components of the full \ecam workflow. More information can be found in Ref.  \cite{rubini2024a_phd} and finer details on the neural network architecture and training procedure can be found in Appendix~\ref{chap:nn_appen}.

\subsubsection{Linear Compression}

The encoder in \cref{fig:autoencoder} represents a coordinate system transformation. To understand this transformation, we start with the vector species conservation equations multiplied by the transformation/projection operator $\mathbf{J}_{\phi} \in \mathbb{R}^{M \times K}$ on both sides:
\begin{equation} \label{eq:compact_species}
   \mathbf{J}_{\phi} \left(\frac{\partial \rho \boldsymbol{Y}}{\partial t}  + \nabla \cdot (\rho \bvec{u} \boldsymbol{Y}) +  \nabla \cdot (\boldsymbol{D}_{\mathrm{eff.}} \nabla \boldsymbol{Y}) \right) = \rho \mathbf{J}_{\phi} \boldsymbol{\dot{\omega}}
\end{equation}
where $\rho$ is density, $\bvec{u}$ is the velocity vector, and $\boldsymbol{D}_{\mathrm{eff.}}= \rho \bar{\boldsymbol{D}} + \frac{\mu_t}{Sc_t} \in \mathbb{R}^{K \times K}$ is a square diagonal matrix of effective diffusion coefficients for all species, and 
\begin{equation} \label{eq:project_species}
    \mathbf{J}_{\phi} = \frac{\partial \boldsymbol{Z}}{\partial \boldsymbol{Y}}  \equiv \bigg[\frac{\partial \boldsymbol{Z}}{\partial Y_1}, \ldots, \frac{\partial \boldsymbol{Z}}{\partial Y_K} \bigg] = f(\text{NN params., activations}),
\end{equation}
represents the Jacobian of the transformation mapping $\phi$ provided by the encoder (see \cref{fig:autoencoder}).

The encoder can either be linear or nonlinear. Within \ecam, a \textit{linear} mapping is used to prevent additional nontrivial closure terms appearing in the latent space governing equations. This means that $\mathbf{J}_{\phi}$ becomes a $M \times K$ linear projection matrix composed of constant weightings that do not vary in space \& time and are independent of thermochemical state. The projection matrix can therefore pass through the differential operators:
\begin{equation} \label{eq:project_species_linear}
    \frac{\partial \rho \boldsymbol{Z}}{\partial t} +  \nabla \cdot (\rho \bvec{u} \boldsymbol{Z}) + \nabla \cdot (\boldsymbol{D}_{Z} \nabla \boldsymbol{Z})  = \rho \textcolor{red}{\boldsymbol{\dot{\omega}}_{Z}} (T, p, \boldsymbol{Z}),
\end{equation}
where $\textcolor{red}{\boldsymbol{\dot{\omega}}_{Z}} (T, p, \boldsymbol{Z}) = \mathbf{J}_{\phi} \boldsymbol{\dot{\omega}}$, $\boldsymbol{Z} = \mathbf{J}_{\phi} \boldsymbol{Y}$ and $\boldsymbol{D}_{Z} \in \mathbb{R}^{M \times M}$ is the diffusivity matrix for the metaspecies. This is equivalent to $\boldsymbol{D}_{Z} = \mathbf{J}_{\phi} \boldsymbol{D}_{\mathrm{eff.}} \mathbf{J}_{\phi}^{-1}$, which simplifies to $\boldsymbol{D}_{Z} \equiv \left(\rho \bar{D} + \frac{\mu_t}{Sc_t}\right)\boldsymbol{I}$, where $\boldsymbol{I} \in \mathbb{R}^{M \times M}$ is the identity matrix, assuming that the molecular diffusion coefficient for all species is equal to $\bar{D}$. Effectively, the latent space metaspecies become linear combinations of the high-dimensional physical species mass fractions, for example:
\begin{equation}
    Z_1 = a_1 Y_1 + a_2 Y_2 + \ldots + a_K Y_K,
\end{equation}
where $a_1$ to $a_K$ are the optimized NN weights (see \cref{fig:autoencoder}). The projection matrix is formed from an autoencoder-like NN with only a single hidden layer and linear activation functions. Although this is approximately equivalent to PCA \cite{Goodfellow-et-al-2016}, when it is formulated as a neural network, the latest specialized ML optimization algorithms can help training scale to very large datasets not possible with standard PCA.

\subsubsection{Non-Linear Reconstruction}

In contrast with the encoder, the decoder is formed from a \textit{nonlinear} multi-layer fully-connected autoencoder NN. The decoder is used to reconstruct the physical species mass fractions from low-dimensional space through the following transformation $\boldsymbol{\widehat{Y}} = \mathbf{J}_{\psi} \boldsymbol{Z}$, where the Jacobian of the decoder $\psi : \mathbb{R}^M \mapsto \mathbb{R}^K$ is $\mathbf{J}_{\psi} = \frac{\partial \boldsymbol{Y}}{\partial \boldsymbol{Z}}$ and $\widehat{\square}$ signifies a reconstructed quantity (see \cref{fig:autoencoder}). Despite using \textit{linear encoding} for the reasons described above, applying \textit{non-linear decoding} can condition the manifold while providing a smoother and higher-quality topology \cite{ZDYBAL2023100859}. In contrast with encoding, nonlinear reconstruction does not introduce additional nontrivial closure terms because it is only the variables that are decompressed, not the governing equations. Although nonlinearity can benefit the representational capacity and manifold topology, it may also have adverse effects on generalization that should be further investigated.

\subsubsection{Latent Space Dynamic Response Regression}

In parallel with learning the low-dimensional manifold, a nonlinear regression mapping $$[\boldsymbol{\dot{\omega}}_{Z}, \dot{\omega}_{T}, c_p, c_v, \bar{\mu}]^{\top} = \mathscr{F}_{\mathrm{NN},Z} (T, p, \boldsymbol{Z})$$ (see \cref{fig:autoencoder}) is learned for the dynamic response of the reaction system projected into latent space (see \cref{eq:project_species_linear}) for a given the state $T, p, \boldsymbol{Z}$. A fully-connected feed-forward neural network---a highly rich and flexible function approximator---is chosen over other methods used in the literature, such as Gaussian processes, which were used by Malik \etal \cite{MALIK20212635}. Neural networks are preferred because of their scalability to a much larger corpus of training data than other methods. Since Gaussian processes and other such nonparametric kernel methods effectively perform inference through a weighted average over \textit{all} points in the training dataset, training and inference times scale poorly ($\mathcal{O}(N^2)$) to large datasets.

It was decided to learn the source terms $\boldsymbol{\dot{\omega}}_{Z}$ (from \cref{eq:compact_species}) rather than the solution propagator $\boldsymbol{Z}(t) \to \boldsymbol{Z}(t + \Delta t)$, as it is simpler to couple with the CFD solver (since operator splitting is not required) . Furthermore, notice that, in \cref{fig:autoencoder}, the NN for the reaction heat absorption rate and the thermophysical properties are also mapped by the function approximator and are now a function of metaspecies rather than physical species mass fractions.

During the development of the \ecam workflow, three key design decisions were made to improve performance. First, it was decided to incorporate the temperature and pressure dependency of directly as inputs to $\mathscr{F}_{\mathrm{NN},Z} (T, p, \boldsymbol{Z})$ mapping (see \cref{fig:autoencoder}), rather than passing $T$ \& $p$ through the encoder layer. This simplifies and accelerates coupling with the flow solver because the thermochemical state does not need to be repeatedly encoded and decoded at every time integration step to follow changes in temperature. Encoding is done once at the start of a simulation and decoding is done once at the end of a simulation.

Second, rather than independently training the NN parameters for the low-dimensional manifold mapping from the parameters for the latent-space dynamic response mapping, they are both trained concurrently. This coupled training procedure helps to avoid complications with a non-unique and non-smooth manifold topology discussed in Ref. \cite{Zdybal2022}. The combined loss function naturally penalizes poor manifold quality as the fitting error for $\mathscr{F}_{\mathrm{NN},Z} (T, p, \boldsymbol{Z})$ implicitly increases.

Finally, in contrast to other researchers who have used approaches physically constrained by Arrhenius's rate law and the law of mass action in latent space \cite{refId0}, we have adopted a purely data-driven approach to approximate $\mathscr{F}_{\mathrm{NN},Z} (T, p, \boldsymbol{Z})$. This is because our early investigations indicated that the representational capacity and the number of trainable parameters of the physics-inspired approach were insufficient, resulting in an intolerably low level of accuracy. For example, with three metaspecies, a physics-inspired latent dynamics approach would only have around 21 parameters to fit, which is not the case for a purely data-driven approach. 

\begin{figure*}[htb]
    \centering\includegraphics[width=183mm]{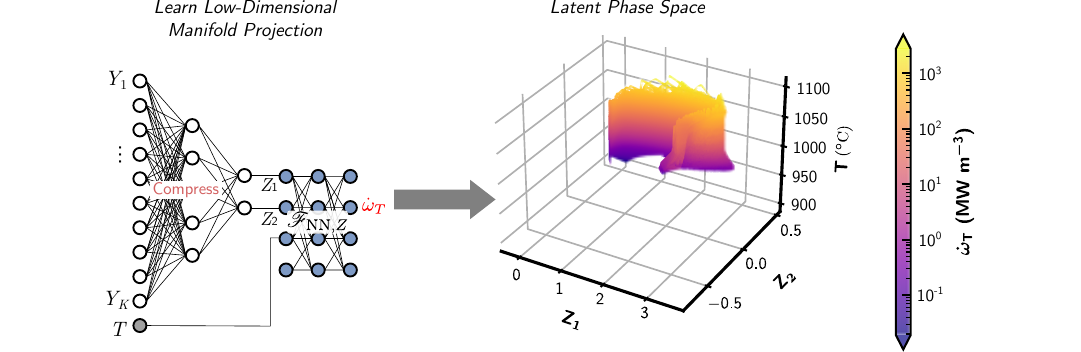} 
    \caption{Latent space manifold for the training database constructed from $Z_1, Z_2$ \& $T$ and colored by the reaction heat absorption rate $\dot{\omega}_T$ as the dependent variable (on a log-scale).}
    \label{fig:latent_phase_space}
\end{figure*}


\subsubsection{Loss Function} \label{sec:loss_func_defi}

The NN model was built and subsequently trained using the \textsc{Keras} \cite{chollet2015keras} and \tensorflow v2.14.0 \cite{tensorflow2015-whitepaper} APIs, in which the loss function to be minimized (\ie $\underset{\boldsymbol{\theta}}{\text{arg\,min } \mathscr{L}}$) is defined as follows $\mathscr{L} = w_1 \mathscr{L}_{\mathrm{chemistry}} + w_2 \mathscr{L}_{\mathrm{thermo}}$, where
\begin{multline} \label{eq:loss_func_all}
    \mathscr{L}_{\mathrm{chemistry}} = \underbrace{\lambda_1 loss \big( \widehat{\boldsymbol{Y}} - \boldsymbol{Y}\big)}_{\text{reconstruction}} + \underbrace{ \lambda_2 loss \big( 
    trans(\boldsymbol{\dot{\omega}}_Z) - trans(\mathbf{J}_{\phi} \boldsymbol{\dot{\omega}}) \big)}_{\phi: \text{ encoded dynamics}}\\
    + \underbrace{ \lambda_3 loss \big( 
    trans(\boldsymbol{\dot{\omega}}) - trans(\mathbf{J}_{\psi} \boldsymbol{\dot{\omega}}_Z) \big)}_{\psi: \text{ decoded dynamics}}\\
\end{multline}
in which $loss(\boldsymbol{Y})$ is defined as $$loss(\boldsymbol{Y}) = \sum_{n=1}^N \sum_{k=1}^{K} \log{(\cosh{(Y_{k,n}}))} / (NK)$$ and $trans(x)$ as  $$trans(x) = \text{sign}(x + 0.0001) \sqrt{|x + 0.0001|}$$ \cite{ZDYBAL2023100859}. The second component of $\mathscr{L}$ is defined as 
\begin{multline} 
    \mathscr{L}_{\mathrm{thermo}} = \text{MSE}(\widehat{c}_p - c_p) + \text{MSE}(\widehat{c}_v - c_v) + \\loss  (
    trans(\widehat{\dot{\omega}}_T) - trans(\dot{\omega}_T)),
\end{multline}
where MSE($\mathbf{X}$) = $\sum_{n=1}^{N} (x_n - \widehat{x}_n)^2 / N$ is the mean square error. The structure of the loss function can be broken down as follows.

\begin{itemize}
    \item \textit{Reconstruction:} the reconstruction loss is the information lost from encoding $\to$ decoding.

    \item \textit{Encoded dynamics}: an error signal is generated between the actual net species production rates projected into latent space and those predicted by $\mathscr{F}_{\mathrm{NN},Z} (T, p, \boldsymbol{Z})$.

    \item \textit{Decoded dynamics}: an error signal is generated between the predicted latent space dynamics projected into physical space and the actual net species production rates.

    \item $trans(\cdot)$: training NNs is challenging when the outputs vary over many orders of magnitude. Since this is the case for $\boldsymbol{\dot{\omega}}$ \& $\dot{\omega}_T$ (see \cref{fig:training_data_profiles}), they are power-transformed to prevent training from skewing toward extreme values by stabilizing the statistical variance \cite{D3RE00212H}. Errors at low orders are as important as those at high orders. This transformation tends to linearize the input-output relationship, making it easier to learn.
  

\begin{figure*}[htb]
    \centering\includegraphics[width=183mm]{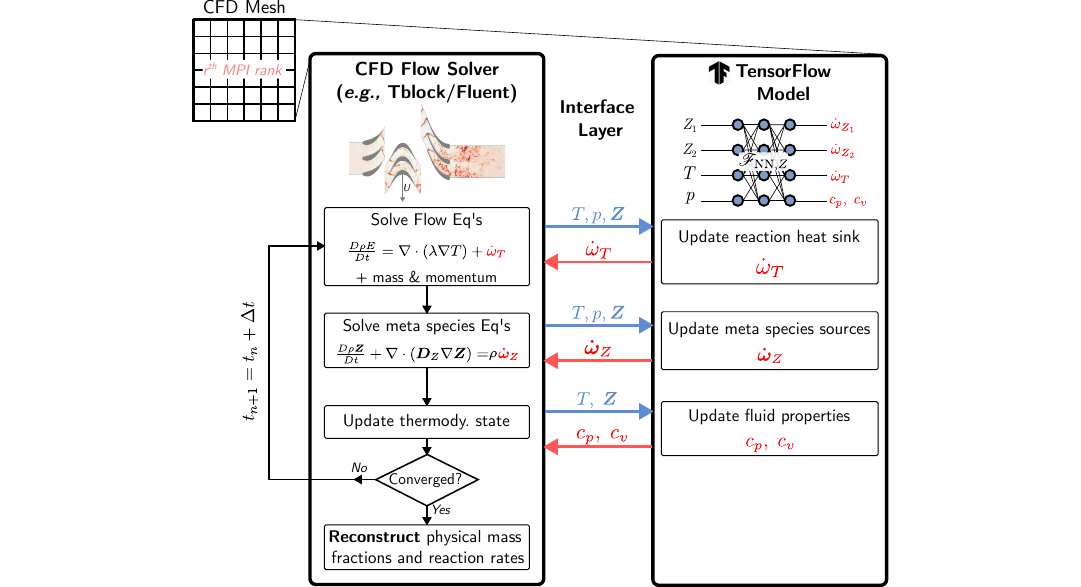} 
    \caption{Flow chart of the coupling process between the CFD flow solver and the trained  \ecam model written using the \tensorflow API.}
    \label{fig:code_coupling}
\end{figure*}

    \item  $loss(\cdot)$: the log-cosh loss function \cite{saleh2024statisticalpropertieslogcoshloss} is used for similar reasons as above. A MSE loss function would bias the model strongly towards high magnitude values for $\boldsymbol{\dot{\omega}}$ \& $\dot{\omega}_T$, whereas log-cosh provides a more balanced treatment across a range of orders of magnitude.
    
\end{itemize}

The multiscale nature of some of the dependent variables is a critical challenge for the learning algorithm \cite{D3RE00212H}. For more details on training the ML model and for values used for the hyperparameters, the reader is referred to Appendix~\ref{chap:nn_appen}. In the current example, training takes approximately 12 to 24 hours.

\subsubsection{Data Preprocessing Pipeline}

There are two main steps to preprocessing before training, culling unimportant species and scaling the inputs and outputs. The first step in the preprocessing pipeline is to eliminate minor species with low variance across the training dataset using \texttt{VarianceThreshold()} from \textsc{Scikit-learn}. These are typically fast-reacting highly nonlinear intermediate radical species whose concentration distributions we are not interested in predicting. This makes training easier, improves the topology of the manifold, and eliminates numerical stiffness without requiring sophisticated treatments. Following this, it is standard practice to normalize the different inputs (\eg temperature and mass fractions) and outputs on the same scale of [0, 1] to balance optimization over the wide range of scales present in the data.


\subsubsection{Latent Phase Space Manifold}

\Cref{fig:latent_phase_space} shows a visualization of the high-dimensional training data projected into low-dimensional latent space via $\boldsymbol{Z} = \mathbf{J}_{\phi} \boldsymbol{Y}$. In this example, the bottleneck layer is limited to 2 latent metaspecies $Z_1$ \& $Z_2$ to enable visualization.  It can be seen that the manifold (consisting of 4 million training data points) is non-overlapping and that the reaction heat absorption rate varies smoothly across it without suffering from issues of non-uniqueness. This indicates robustness, underlining the benefits of simultaneously learning the manifold and the non-linear dynamics regression. 


\begin{figure*}[htb]
    \centering\includegraphics[width=183mm]{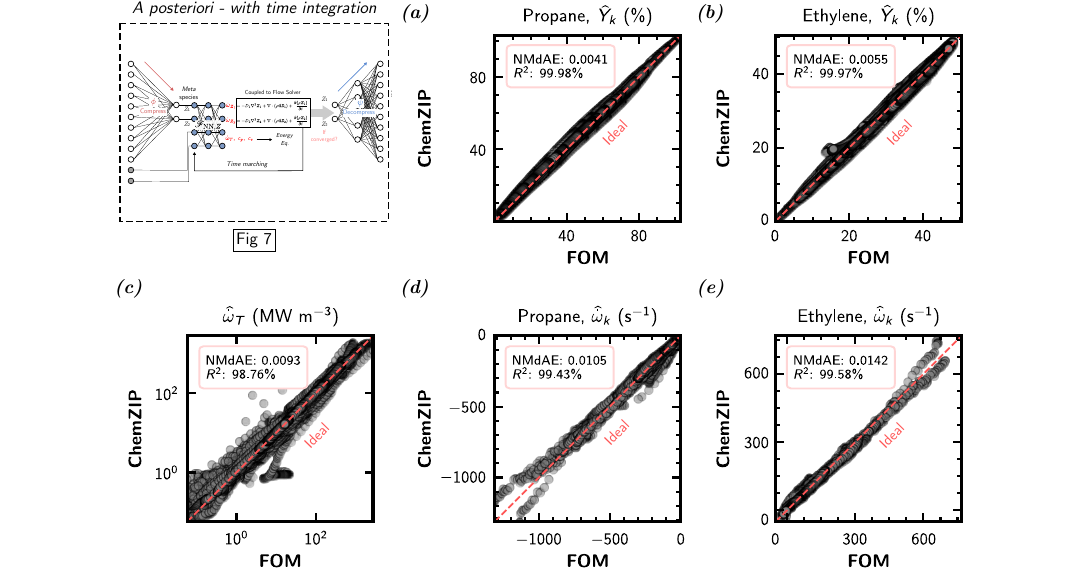} 
    \caption{A posteriori (\ie \textit{with} time integration) parity plot (500,000 data points) comparing reconstructed \ecam predictions against the ground truth (FOM) data for \textbf{(\textit{a})}--\textbf{(\textit{b})} the propane/ethylene dry yields \textbf{(\textit{c})} the heat absorption rate (on a log scale) and \textbf{(\textit{f})}--\textbf{(\textit{e})} the propane/ethylene production rates.}
    \label{fig:parity_time_integrated}
\end{figure*}

\subsection{Coupling \chemzip with Flow Solver}

The trained \ecam model written using \tensorflow (TF) can be deployed into any CFD flow solver relatively easily by executing the steps summarized below and illustrated in \cref{fig:code_coupling}.
\begin{enumerate}

    \item For each of the $M$ metaspecies (where $M \sim 2 - 4$), add a convection-diffusion transport equation of the form shown in \cref{eq:project_species_linear} using built-in discretization methods.\\

    \item To update the source terms $\textcolor{red}{\boldsymbol{\dot{\omega}}_Z}$  at each time iteration given new values for $T, p, \boldsymbol{Z}$, the TF model for $\mathscr{F}_{\mathrm{NN},Z} (T, p, \boldsymbol{Z})$ must be queried in each mesh cell.\\

    \item Add metaspecies wall boundary condition, which is a zero diffusive flux condition projected onto the latent space, \ie $\left. \mathbf{J}_{\phi} \nabla \boldsymbol{Y} \right|_{\text{wall}} \equiv  \left. \nabla \boldsymbol{Z} \right|_{\text{wall}} = 0$ \cite{anderson1989hypersonic}.\\

    \item Add additional heat absorption sink term $\textcolor{red}{\dot{\omega}_T}$ to the energy equation. This source term is the dominant mechanism driving interactions between the aerothermodynamics and the chemistry. \\

    \item Clip the inputs to the NNs within known ranges set by the training space to prevent dangerous model extrapolation.\\

    \item Properties $\textcolor{red}{c_p}$ \& $\textcolor{red}{c_v}$ are now set as a function of $T$ \& $\boldsymbol{Z}$.
    
\end{enumerate}
Effectively, all complex reaction effects are lumped into a set of source terms and changes in fluid properties. Crucially, as will be demonstrated in \Cref{sec:results}, the stiffness of the reaction system is alleviated because high-frequency species are filtered by the encoder. This means that the latent chemistry does not impose any time-step restrictions on the datum flow solver. Therefore, the basic \textit{explicit} time integration scheme, multigrid acceleration, etc., can be used without modification, significantly accelerating convergence. Importantly, the coupling interface between the flow solver and \tensorflow imposes negligible overhead.

This methodology is agnostic to the level of solver type (\eg \tblock or \fluent) and its fidelity, which could be anything from RANS to LES. Currently, \ecam is coupled with the industry-standard turbomachinery flow solver \tblock \cite{10.1115/1.2927983}.

\section{Verification and Benchmarking} \label{sec:results}
This section presents a verification and performance benchmarking assessment of \chemzip. To evaluate its accuracy, robustness, and generalizability, it is compared to the full-order model (FOM) for both a 1D fully-mixed PFR model as well as 3D viscous reacting flow simulations from a database of unseen test conditions.

Throughout this chapter, the kinetic model used is a 65-species propane--steam pyrolysis mechanism typical of industrial steam cracking, which is generated automatically using \textsc{Rmg-Py} \cite{GAO2016212}.  For now, the mechanism is small to accelerate the development time; however, an arbitrarily complex mechanism can be used as a drop-in replacement when \chemzip is deployed in practice.


\begin{figure*}[htb]
    \centering\includegraphics[width=183mm]{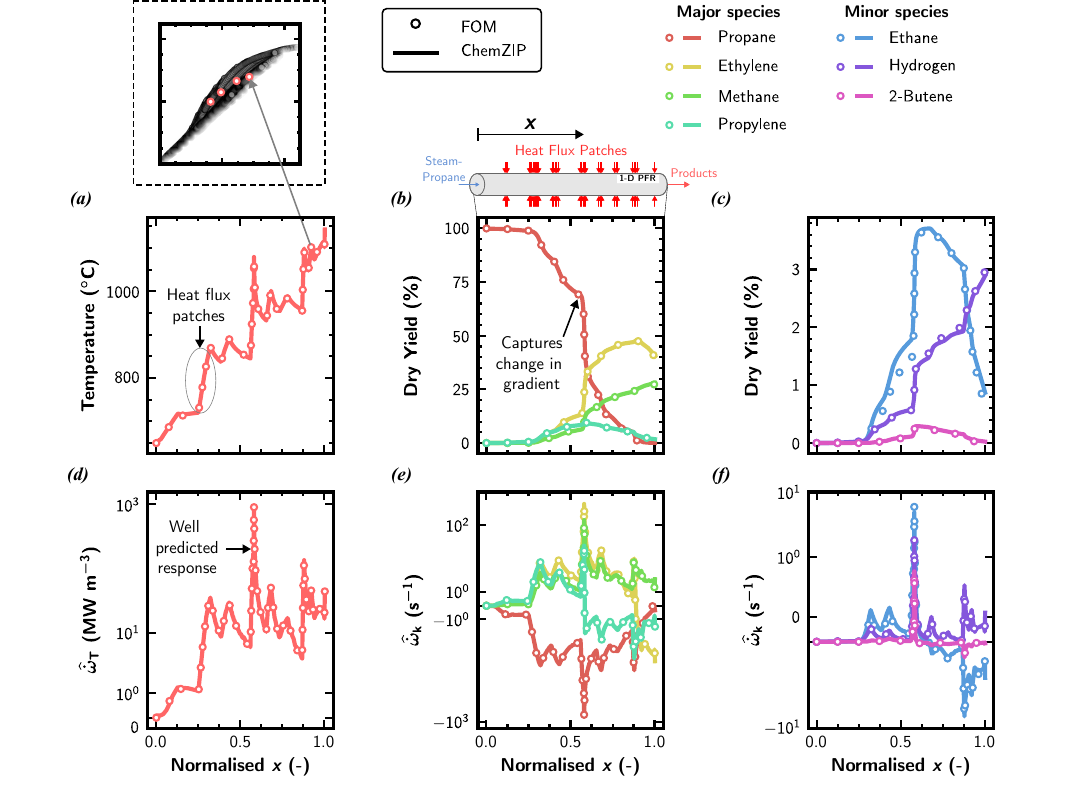} 
    \caption{1D PFR time integrated solution history for \textbf{(\textit{a})} temperature \textbf{(\textit{b})}--\textbf{(\textit{c})} major/minor species dry yields \textbf{(\textit{d})} the reaction heat absorption rate and \textbf{(\textit{e})}--\textbf{(\textit{f})} the  major/minor species net production rates. Note that the circular markers denoting the FOM are significantly downsampled for clarity.}
    \label{fig:time_integrated_species}
\end{figure*}

\subsection{Errors Metrics}

To assess the accuracy and generalisability of the trained \ecam model on $N_{\mathrm{test}}$ unseen test data samples, the $R^2$ score, the normalized median absolute error (NMdAE) and the normalized root mean square error (NRMSE) metrics are used to measure goodness of fit. Taking the reconstruction of the $k$\th species mass fraction as an example, these three metrics are defined as
\begin{equation} \label{eq:r2_score}
    R^2 = \left(1 - \frac{\sum_n^{N_{\mathrm{test}}} (Y_{k,n} - \widehat{Y}_{k,n})^2}{\sum_n^{N_{\mathrm{test}}} (Y_{k,n} - \overline{Y}_{k})^2} \right),
\end{equation}
\begin{equation} \label{eq:nmdae}
    \mathrm{NMdAE} = \frac{\text{Median}\left(\left | Y_{k,1} - \widehat{Y}_{k,1} \right |, \ldots, \left | Y_{k,N_{\mathrm{test}}} - \widehat{Y}_{k,N_{\mathrm{test}}} \right |   \right)}{\left | \overline{Y}_k \right |},
\end{equation}
and
\begin{equation} \label{eq:nrmse}
    \mathrm{NRMSE} =  \frac{ \sqrt{\sum_n^{N_{\mathrm{test}}}(Y_{k,n} -\widehat{Y}_{k,n})^2}}{  \left | \overline{Y}_k \right |  \cdot N_{\mathrm{test}}}
\end{equation}
where $Y_{k,n}$ is the ground truth, $\widehat{Y}_{k,n}$ is the prediction for the $k$\th species \& $n$\th sample, and $\overline{Y}_{k} = \frac{1}{N_{\mathrm{test}}} \sum_n^{N_\mathrm{test}}Y_{k,n}$. Note that the $R^2$ metric ranges between 0\% and 100\%, where 100\% represents a perfect prediction.


\subsection{Verification A Posteriori: Multistep Fully-Mixed 1D PFR}

The performance of \chemzip is assessed in a time-integrated 1D PFR environment in order to assess not only (a priori) input--output performance, but also (a posteriori) sensitivity to error accumulation. The test dataset consists of 500,000 data points corresponding to approximately 1000 different complex heat flux profile realizations withheld from the training dataset. The trained model is integrated using an explicit 4\th/5\th-order adaptive Runge-Kutta method. The very fact that an \textit{explicit} method works (rather than a stiff \textit{implicit} method), implies that the numerical stiffness of the reaction system has been alleviated, dramatically accelerating convergence. This will be elaborated on in \cref{sec:benchmark}.

\Cref{fig:time_integrated_species} shows the solution history for a single heat flux profile realization while \cref{fig:parity_time_integrated} condenses the time histories for all 1000 1D PFR simulations from the test dataset into a series of parity plots. Assessing the performance of \ecam when subjected to sharp gradients in temperature, as seen in \cref{fig:time_integrated_species}, ensures representative conditions for the turbo-reactor are covered. It is clear that even for a complex external heat load (see \cref{fig:time_integrated_species}(\textit{a})), accurate production rates can be reconstructed from latent space (see \cref{fig:time_integrated_species}(\textit{d}))--(\textit{f})). As a result, following a heat flux perturbation, the temporal gradient in the yield responds accurately throughout the entire evolution for both major and minor species. For the major species, the predictions are almost perfect. For lower concentration species, \eg ethane, the accuracy is lower (see \cref{fig:time_integrated_species}(\textit{c})), as expected. However, interestingly, it appears that the prediction errors are damped over the evolution and that the ethane yield at the outlet is, in fact, well predicted.

To determine whether the conclusions drawn above generalize to all time histories, the complete integrated test database is plotted in \cref{fig:parity_time_integrated}. Overall, the $R^2$ score is above 95\% for all quantities, suggesting good performance. The accuracy of the yields is higher than that of the production rates, which indicates that errors in the production rates are damped by the integration operator, which is favorable because the yields are the quantities of interest to the reaction system designer.

\begin{figure}[htb]
    \centering\includegraphics[width=88mm]{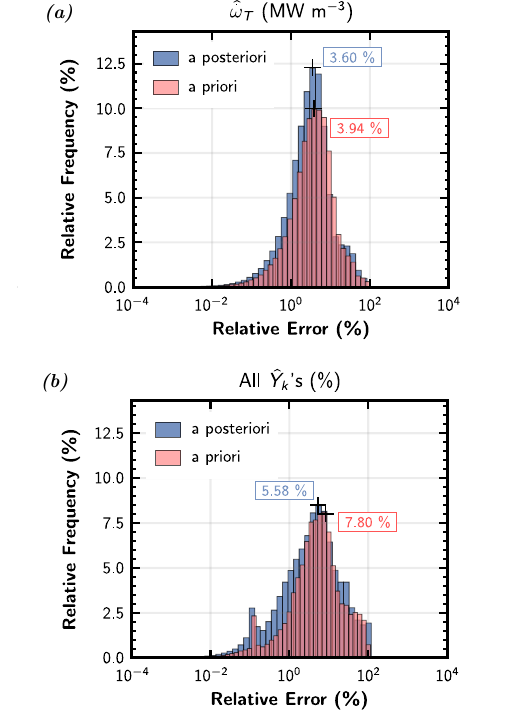} 
    \caption{Histograms for the relative error ($x$) = $\left| (x - \widehat{x}) / x \right|$  between \ecam  and the ground truth FOM for time-integrated (a posteriori) and non-time-integrated (a priori) test data for \textbf{(\textit{a})} the reaction heat absorption rate and \textbf{(\textit{b})} all the species mass fractions. The abscissa is on a log scale.}
    \label{fig:predictions_histogram}
\end{figure}

\Cref{fig:predictions_histogram} shows the distribution of prediction errors over 500,000 test points for a priori (single-step predictions) and a posteriori (multistep time integrated) predictions. Both histograms are centered on a relative error of less than \textbf{10\%}.  There is no significant difference between the a priori and a posterior error distributions, suggesting a minimal sensitivity of the model to time-integrated error accumulation. A possible explanation for this robustness is the inherent self-correcting nature of an endothermic dynamical system. If the temperature is too high due to modeling errors, the reaction heat absorption increases and brings down the temperature toward the ground truth; conversely, if the temperature is too low, the reaction heat absorption decreases and brings the temperature back up. This physical mechanism suppresses the growth of errors and encourages robustness.

\begin{figure}[htb]
    \centering\includegraphics[width=88mm]{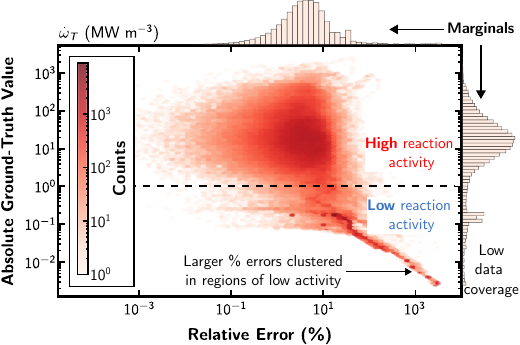} 
    \caption{2D joint and marginal distributions for the relative prediction errors and absolute value for the reaction heat absorption rate (500,000 test data points). Both axes are on a log scale.}
    \label{fig:histbin}
\end{figure}

However, in both cases, \cref{fig:predictions_histogram} shows a small fraction of relatively high percentage errors ($(x - \widehat{x})/x$). This is primarily due to poor numerical conditioning and bias of this metric as the values of the ground-truth in the denominator become small. This effect is most pronounced for the mass fractions, which are often precisely equal to zero. However, \cref{fig:parity_time_integrated} and other investigations (not shown) indicate that a poor relative error close to zero does not affect the global solution. Furthermore, focusing on the reaction heat absorption rate as it represents an integrated effect of dynamics in all species,  \cref{fig:histbin} suggests that generally larger relative errors occur in regions of low reaction activity. This is because of limited training data coverage in this low temperature regime. The impact on the yield should be small because the predicted values are still below or close to the activation threshold of the reaction, and the model can be manually turned off in this regime.

Despite this, more generally, the analysis presented provides valuable information that can be fed back into the training data generation process. Given these insights, the modeler can cluster more training samples in underrepresented regions to counteract this bias if necessary.

Ensuring consistent accuracy across several orders of magnitude for reaction responses and species yields underscores perhaps the principal challenge in optimizing a neural network for chemical kinetics: the inherent multiscale dynamics of the system. This leads to various optimization challenges summarized as follows: (1) the optimizer tends to focus on certain scales (2) large variability in scales within a minibatch can result in difficult learning dynamics and instability in the gradient descent optimization algorithm and (3) the predictions are highly sensitive to hyperparameter selection, which may be optimum for some scales but not for others. Some of these challenges are remedied by choosing an appropriate loss function, as described in \cref{sec:loss_func_defi}; however, further refinement is still needed.


\subsection{3D Heated Duct Flow}

To verify the robustness and generalizability of \ecam to a more complex ``out-of-distribution'' flow environment, the model is deployed into a full-fledged CFD solver for a 3-D heated duct test case. This multidimensional viscous domain is entirely different from the training dataset. \ecam is benchmarked and verified against an industry-standard commercial reacting flow solver \fluent with a detailed kinetic model (\ie the FOM). For performance benchmarking, \ecam is also compared against state-of-the-art acceleration strategies available in \fluent, that is, a combination of DACM (see \cref{sec:physics_based}) + dynamic cell zoning (DCZ) \cite{doi:10.1080/00102200903190836}, which will be denoted as ``Fluent Accel.'' in later figures.

\subsubsection{Domain and Boundary Conditions}

\begin{figure}[htb]
    \centering\includegraphics[width=88mm]{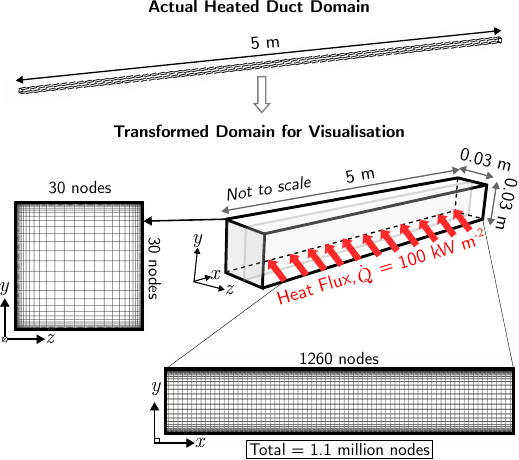} 
    \caption{Heated duct test case domain and computational mesh (totaling 1.1 million nodes).}
    \label{fig:domain_mesh}
\end{figure}

\Cref{fig:domain_mesh} shows the computational domain and the mesh. The dimensions of the duct and the boundary condition (see \Cref{tab:BCs}) are are representative of a section of a typical steam cracking coil. The heat input $\dot{Q}$ is imposed as a Neumann boundary condition uniformly for the four walls. Employing a long domain (\ie \SI{5}{\meter}) is necessary to ensure the feed undergoes substantial transformation, which can be a slow process due to thermal diffusion limitations of heat transfer. To improve visualization, contour plots throughout this chapter will be presented using an axially-squashed domain, as shown in \cref{fig:domain_mesh}. 

The same mesh is used for the simulations in \fluent and in \ecam. The maximum non-dimensional wall distance in the domain is $y^+ = 7$; therefore, adaptive velocity wall functions are employed with the standard Spalart-Allmaras turbulence model in both solvers. Unless stated otherwise, simulations can be assumed to be performed in parallel on a high-performance Intel Cascade Lake 18-core CPU.

\begin{table}[htb]
\centering
\caption{Boundary conditions for heated duct}\label{tab:BCs}
\begin{tabular}{lc} 
\toprule
\toprule

\textbf{Boundary Conditions} &   \\
\midrule
Inlet total pressure (bar) & 2.1 \\
Exit static pressure (bar) & 2.0 \\
Inlet total temperature (\si{\celsius}) & 650 \\
Wall heat flux, $\dot{Q}$ (\si{\kilo\watt\per\meter\squared}) & 100 \& 150 \\
Inlet turbulent to laminar viscosity ratio & 10 \\
\bottomrule
\bottomrule
\end{tabular}
\end{table}


\subsubsection{Computational Performance Benchmarking}  \label{sec:benchmark}

\begin{figure}[htb]
    \centering\includegraphics[width=88mm]{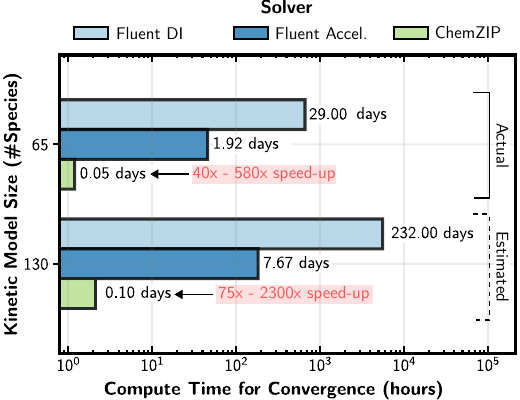} 
    \caption{Computational acceleration achieved by \ecam compared with direct integration (DI) in \fluent and a state-of-the-art acceleration (Accel.) strategy that combines DACM + DCZ.}
    \label{fig:speedup}
\end{figure}

Performance benchmarking is achieved by measuring the wall clock time required to march the solution to a steady state for (a) the FOM in \fluent (b) an accelerated scheme for the chemistry and (c) \ecam. \Cref{fig:speedup} demonstrates that significant computational acceleration can be achieved using \ecam. For a 65-species mechanism, a speedup of \textbf{580$\bm{\times}$} is measured compared to direct integration and \textbf{40$\bm{\times}$} compared to state-of-the-art acceleration strategies. Based on scaling rules, it is estimated that for a more detailed (and more accurate) mechanism, the rewards are even more substantial, with \ecam potentially capable of at least a \textbf{75$\bm{\times}$} speedup over ``Fluent Accel.'' (see \cref{fig:speedup}). Furthermore, greater speedups are possible for larger mesh sizes.

\begin{figure}[htb]
    \centering\includegraphics[width=88mm]{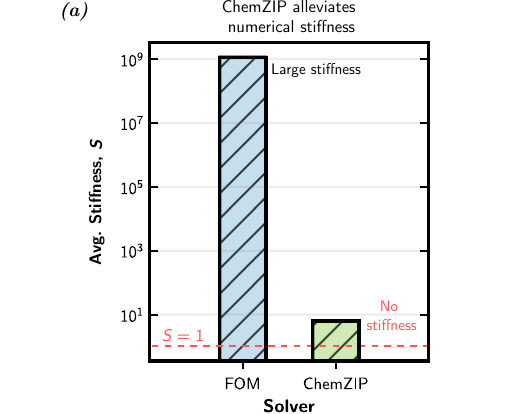} 
    \caption{For a 65-species propane mechanism: \textbf{(\textit{a})} a comparison of the average numerical stiffness for the chemistry (on a log scale).}
    \label{fig:stiffness_ode}
\end{figure}

\begin{figure}[htb]
    \centering\includegraphics[width=88mm]{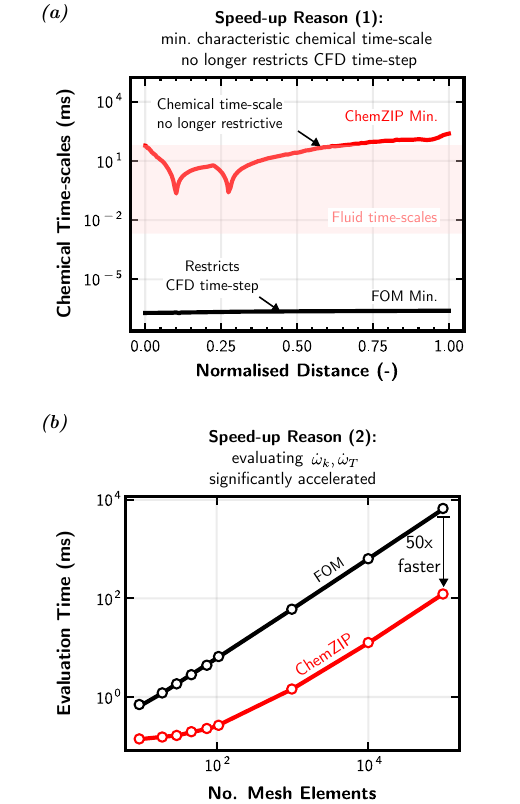} 
    \caption{For a 65-species propane mechanism: \textbf{(\textit{a})} a comparison of the streamwise variation in minimum characteristic chemical time-scale between \ecam and the FOM (on a log scale) and \textbf{(\textit{b})} comparison of single-core CPU time to evaluate the chemical and energy source terms as a function of number of mesh cells in a computational batch (on a log-log scale).}
    \label{fig:chemzip_fom_timescale}
\end{figure}

\Cref{fig:stiffness_ode,fig:chemzip_fom_timescale} illustrate several of the primary contributions to the computational acceleration achieved by \ecam. First, the number of transport equations is reduced, in this example, from 65 to 3. Second, the average numerical stiffness of the chemical system is decreased by a factor of approximately \textbf{$10^9$} (see \cref{fig:stiffness_ode}(\textit{a})). The stiffness is defined as the ratio of the maximum to minimum characteristic chemical time scales $\frac{\text{max}(\boldsymbol{\tau}_{\mathrm{chem.}})}{\text{min}(\boldsymbol{\tau}_{\mathrm{chem.}})}$, where $\boldsymbol{\tau}_{\mathrm{chem.}}$ is calculated from the eigenvalues of the Jacobian matrix $\mathbf{J}$ in physical space or the Jacobian matrix $\mathbf{J}_{Z} =  \frac{\partial \boldsymbol{\dot{\omega}}_{Z}}{\partial \boldsymbol{Z}}$ in latent space \cite{doi:10.1080/00102202.2020.1760257}. It is noted that very long time scales $\tau_{\mathrm{chem.}} > \SI{1000}{\second}$ are filtered out, as they are associated with approximately nonreacting species. It is this intense stiffness of the system that necessitates the use of specialized implicit stiff ODE solvers. These require frequent evaluation and inversion of the Jacobian matrix in each mesh cell, effectively, to scale the solution update appropriately for each species. The Jacobian inversion is the reason why the cost of DI grows cubically with the number of species. 

For the latent dynamical system, fast-reacting intermediate species are filtered out, and thus the stiffness is almost eliminated. Not only does this dramatically improve the rate of convergence to a steady-state solution, but it also enables the use of fast explicit time integrators instead. These do not require the Jacobian matrix and incur a much smaller overhead per step. For the same error tolerances, a factor of 2 performance benefit is possible by integrating the latent system with an explicit method over a stiff implicit one. Furthermore, \cref{fig:chemzip_fom_timescale}(\textit{a}) shows that, unlike the FOM, the latent space system does not impose a restriction on the time-step set by the datum nonreacting flow system. This is because the characteristic chemical time-scale is similar to that of the fluid. Therefore, the baseline numerical methods of the nonreacting CFD solver can be used efficiently and without modification.

Finally, \cref{fig:chemzip_fom_timescale}(\textit{b}) illustrates that for more than 100 mesh cells, the cost of evaluating the chemical and energy source terms is \textbf{50$\bm{\times}$} faster using a NN in latent space compared to the FOM. For batches of less than 100 cells, the latency associated with the \tensorflow API overshadows the performance. However, for large batches of queries to the NN, this cost is amortized, and the scaling becomes linear (see \cref{fig:chemzip_fom_timescale}(\textit{b})). 

The culmination of these factors results in the level of computational acceleration observed previously in \cref{fig:speedup}. Consequently, for the first time, we have demonstrated that the degree of computational acceleration necessary to bring detailed aerochemical modeling into the design optimization loop is now possible. The next question is whether the predictive accuracy is sufficient to perform realistic optimization. 

\begin{figure*}[htbp]
    \centering\includegraphics[width=183mm]{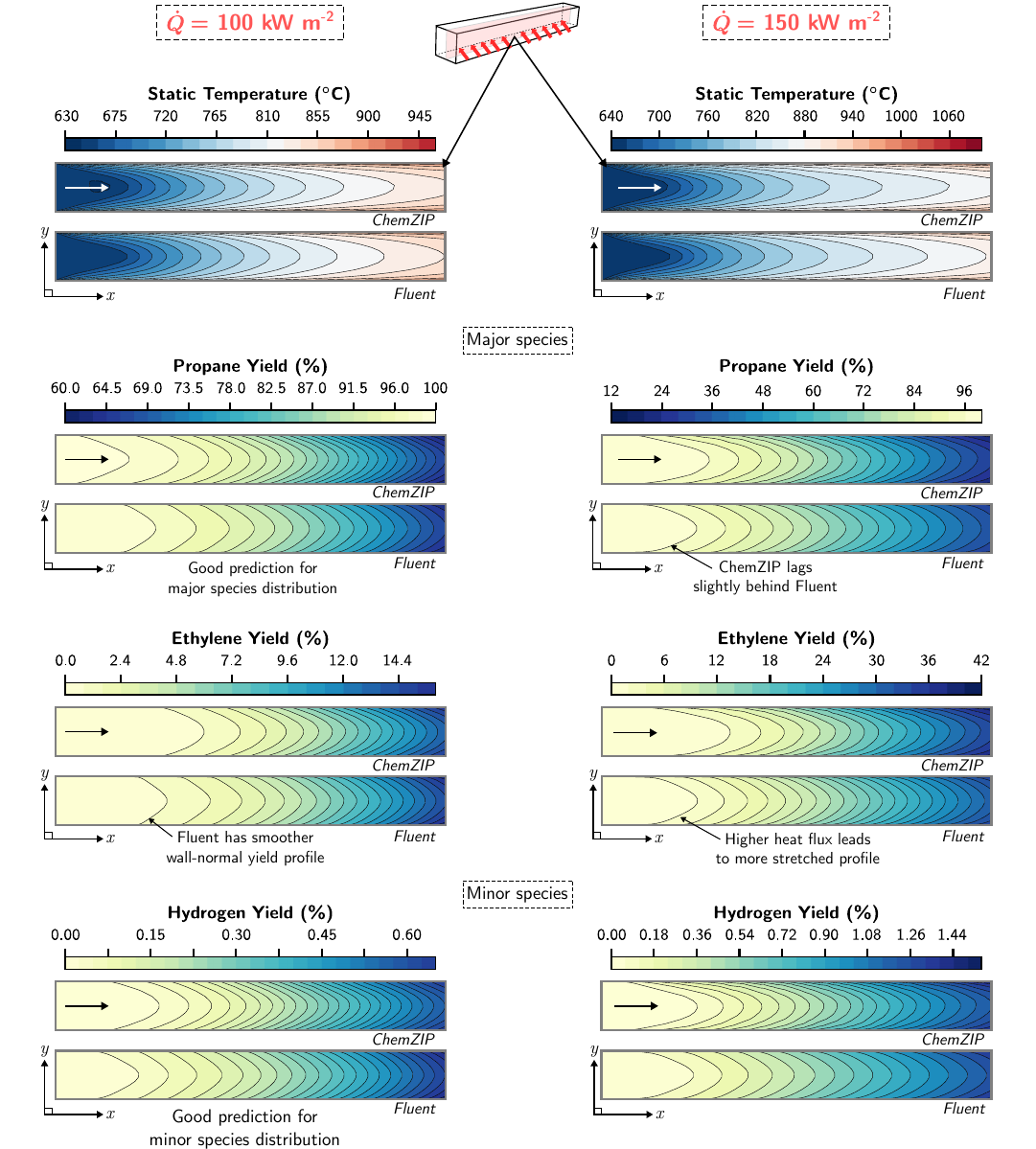} 
    \caption{Verification of the reconstructed temperature and dry yields under two heat loads.}
    \label{fig:yield_compare_pipe}
\end{figure*}

\begin{figure*}[htb!]
    \centering\includegraphics[width=183mm]{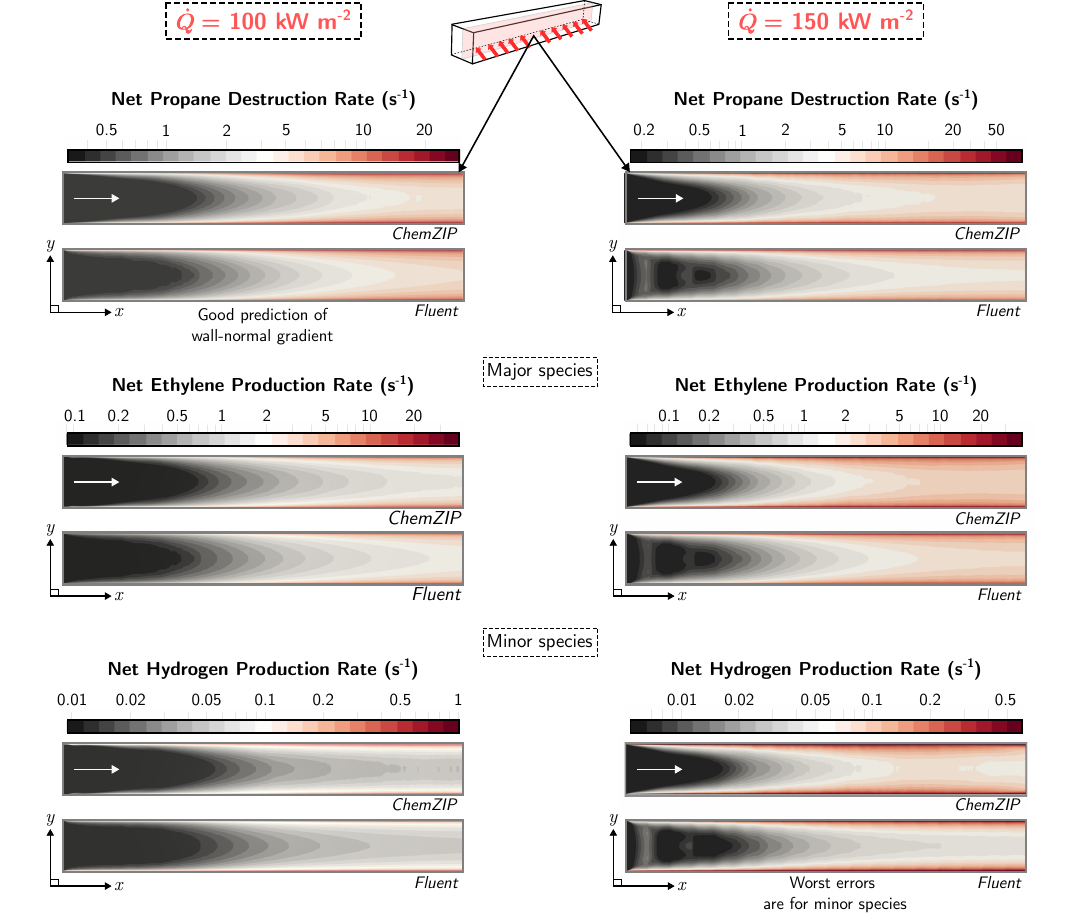} 
    \caption{Verification of the reconstructed the reaction source terms under two heat loads.}
    \label{fig:wdot_compare_pipe2}
\end{figure*}

\subsubsection{Predictive Accuracy}

The accuracy of \ecam is evaluated under two heat flux conditions $\dot{Q} = \SI{100}{\kilo\watt\per\meter\squared}$ and $\dot{Q} = \SI{150}{\kilo\watt\per\meter\squared}$ to exercise its performance in a wider region of the thermochemical space. Although this multidimensional viscous flow environment has not been observed during training, \cref{fig:yield_compare_pipe,fig:wdot_compare_pipe2} shows that subject to standard modeling assumptions for reacting flow \cite{poinsot2005theoretical}, the yields and reaction rates are within several percent of \fluent. Performance is good for both major species, minor species and temperature in both the streamwise direction and the wall-normal direction, and the contours are surprisingly smooth and symmetric. The channel is long enough (\ie \SI{5}{\metre}) that if compounding of model errors was an issue, it would be observed in the results. This does not appear to be a serious problem, as evidenced by a similar streamwise distribution for the yield and reaction rate between the solvers (see \cref{fig:yield_compare_pipe,fig:wdot_compare_pipe2}). However, due to inconsistencies with the underlying numerics of the two solvers, the streamwise yield is slightly ahead of \fluent. This is because in \tblock, the axial velocity is higher and the reaction time-scale is lower. For the higher heat flux condition, the subtle stretching of the yield profile is captured by \ecam, albeit marginally overstretched. 

Inspecting the yields in \cref{fig:yield_compare_pipe}, there is a small discrepancy in the shape of the wall-normal profile. This difference could be attributed in part to the inconsistency between the underlying numerics of the two solvers. An additional explanation could also be that the reaction system does not respond accurately to turbulent diffusion of the species, which is an approximately temperature-independent process and is not incorporated during training. The diffusion term in the FOM produces a gentler wall-normal yield gradient (see \cref{fig:yield_compare_pipe}) due to its smoothing effect. In fact, it is possible that the success of this methodology can perhaps be attributed to the temperature dominance of these types of chemical reaction processes. This is in contrast to other types of reaction processes, such as nonpremixed combustion, where diffusion and mixing can become the limiting physical process. On the other hand, while it is true that the reduced-order system may not react perfectly to species diffusion, it is reassuring that diffusion still has a notable effect of the system dynamics despite this mechanism not being included during training data generation. Despite some inaccuracies seen, the reader is reminded that the objective of the \chemzip surrogate model is to accelerate thermochemical modeling sufficiently so that design optimization is possible. Loss of absolute accuracy can be tolerated if trends are well predicted.


\begin{figure}[htb]
    \centering\includegraphics[width=88mm]{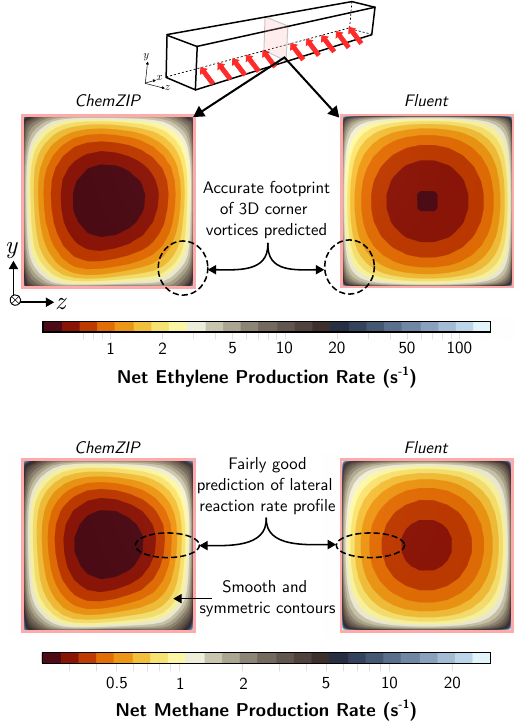} 
    \caption{Net species production rate distributions on an z-y plane cross section ($\dot{Q} =$ \SI{100}{\kilo\watt\per\meter\squared}).}
    \label{fig:wdots_midx}
\end{figure}

For capturing two- and three-dimensional effects not seen before in this chapter, \ecam works notably well. Due to lateral temperature gradients, mass diffusion, and growth of the fluid boundary layer, a nonuniform reaction rate profile is produced from the core to the wall. In addition, secondary vortices lead to enhanced heat transfer and an accumulation of energy at the corners of the channel (see \cref{fig:wdots_midx}). These 2D/3D effects are mostly well captured by \ecam (see \cref{fig:wdot_compare_pipe2,fig:wdots_midx}). Remarkably, the reconstructed production rate source terms respond accurately to wall-normal temperature gradients, which is a challenging task due to the high degree of nonlinearity and sensitivity of the dynamics. 

\begin{figure}[htb]
    \centering\includegraphics[width=88mm]{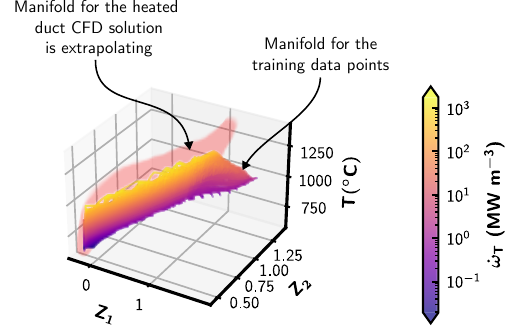} 
    \caption{A projection of the latent phase space manifolds for all points in the training database and for all solution points from the heated duct ($\dot{Q} =$ \SI{150}{\kilo\watt\per\meter\squared}) CFD simulation (shown in red).}
    \label{fig:laten_phase_cfd}
\end{figure}

\Cref{fig:laten_phase_cfd} shows the CFD solution in latent space for all points in the mesh (shown as a red point cloud) superimposed onto the phase space containing the training data manifold. As eluded to previously, due to mass diffusion and two-/three-dimensional effects not captured during training, the regions of the phase space accessed during the CFD computation are slightly outside the training data manifold. This indicates extrapolation. Nevertheless, we have shown that the accuracy is acceptable and the solution is stable, providing confidence in the generalizability of the model.

This section has shown that, in addition to the successful performance benchmarking, the accuracy of \ecam is within reasonable tolerances. Despite the fact that the model was deployed in a very different environment from the 1D training simulator, the spatial profiles of the chemistry were well predicted. This was enabled by intelligent curation of the training database. Since chemical source terms are simply an input--output mapping $\mathscr{F}_{\mathrm{NN},Z} : (T, p, \boldsymbol{Z}) \mapsto \boldsymbol{\dot{\omega}}_{Z}$, training has been performed in such a way as to sample regions of the thermochemical space accessed in more complex flows. 

In the following section, \ecam will be used to solve problems for which it was originally intended: modeling complex aerochemical interactions in the turbo-reactor.

\subsection{Multi-Stage Turbo-Reactor Proof of Concept}

As a proof of concept, the new methodology for accelerated aerochemical coupling is now deployed into the turbo-reactor numerical modeling system. The domain is a multistage turbo-reactor architecture with an inlet temperature of \SI{840}{\celsius}, and the CFD solver is run in transient mode. The loading coefficient of the rotor row is approximately $\Psi = 3.0$. This is a uniquely complex aerodynamic environment with rich aerochemical interactions that were not possible to model before without impractically large computational resources. This provides new insights into the aerochemical flow physics of the turbo-reactor that will inform design optimization for a more efficient chemical reaction in future iterations.

\begin{figure*}[htb]
    \centering\includegraphics[width=183mm]{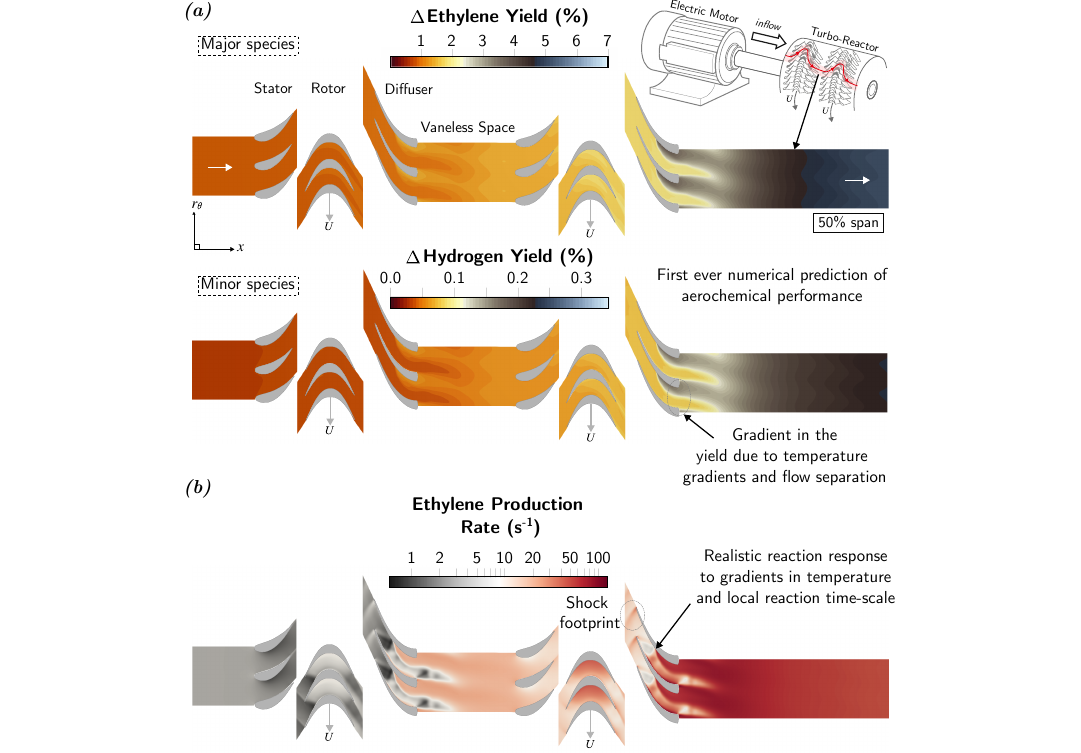} 
    \caption{Time-averaged \textbf{(\textit{a})} change in major and minor species yields relative to the inlet and \textbf{(\textit{b})} ethylene net production rates (on log scale) at midspan across a two-stage turbo-reactor.}
    \label{fig:turboreactor_yield_plots}
\end{figure*}

\Cref{fig:turboreactor_yield_plots}(\textit{a}) shows the time-averaged yield distribution for both major and minor species yields across two stages. Within the second stage diffuser blade passage, the footprint of separated suction surface flow is clearly visible through a pitchwise gradient in the yields. Therefore, the reaction is responding reasonably to regions with different reaction time-scales and temperature histories. The net ethylene production rate shown in \cref{fig:turboreactor_yield_plots}(\textit{b}) also appears to produce realistic responses to temperature gradients. For example, the footprints of several shockwaves (which result in a sharp rise in temperature and hence nonequilibrium response of the reaction) are clearly visible at the diffuser leading edge. 

At this point, due to prohibitive computational costs, it is beyond the scope of this work to verify the accuracy of \ecam for these types of highly complex flows. However, what these results show is that the model provides remarkably realistic and stable solutions despite being deployed in an entirely unfamiliar aerodynamic environment.

\section{Summary and Conclusions} \label{chap:nn_appen}

In this paper, we have comprehensively established the potential of using a machine-learning-enhanced surrogate model to accelerate aerothermochemical modeling within complex engineering systems without significantly sacrificing accuracy. Our approach simultaneously addresses three key challenges of modeling reactive flows by (1) reducing the number of transport equations (2) reducing the cost of evaluating the thermochemical source terms and (3) curtailing numerical stiffness. Not only is the cost per CFD time step reduced, but the steady-state convergence rate is also improved using fast explicit time integration. A key feature of our methodology is bypassing the well-known bottleneck of training data generation, with database development times as low as a few hours.

The \chemzip methodology is 50 times faster than traditional computational acceleration techniques (\eg DZM + DACM) while being substantially more universal than FGM used in combustion modeling. For more accurate mechanisms on large grids, it is likely that the measured speedups will be even higher, of the order of 100$\times$ to 1000$\times$. A key motivation for adopting a neural-network-based data-driven approach is the ability to scale to massive datasets with a large number of species. In the future, this will allow the dynamic response of highly detailed (and thus accurate) kinetic models covering a very wide range of conditions to be embedded and compressed into a quickly accessible mapping. To accomplish this accurately, our approach is to intelligently exercise the complex chemical system by perturbing it within a known envelope of realistic temperature gradients.

A sequence of verification investigations have been conducted to assess accuracy and the findings of these experiments can be summarized as follows:
\begin{itemize}

    \item The $R^2$ metric was above 95\% for all quantities of interest in the 500,000 time-integrated multistep test dataset.
    
    \item The a priori and a posteriori error distributions were remarkably similar, indicating robustness to error accumulation.

    \item In an entirely unfamiliar multidimensional viscous flow environment, \chemzip performed surprisingly well and responded to 3D and diffusion effects despite not being introduced in the training data generation simulations.

    \item The model was deployed in a complex turbomachinery environment for the first time and produced realistic yield and reaction rate distributions.
    
\end{itemize}
So far, turbulence--chemistry interactions and the cross-species heat flux term \cite{rubini2024a_phd} have been neglected. In future work, these effects will be incorporated into the training data generation process to improve the generalizability of \chemzip to more complex flows.

The \chemzip methodology introduced in this paper is ideally suited for design optimization investigations for complex engineering systems, such as the turbo-reactor. In these cases some loss of absolute accuracy can be tolerated assuming trends are well predicted. This step change in aerothermochemical numerical modeling that enables fast iterative design exploration could be a game changer for optimizing the dynamics of reaction systems early in the design cycle. Given the novel problem of modeling highly complex aerothermochemical interactions within turbomachines, strong collaboration between fluid dynamicist and chemical engineers is becoming increasingly important in solving this complex challenge.



\appendix

\section{Neural Network Architecture}


\Crefrange{tab:detailed_network_params1}{tab:detailed_network_params3} shows details of the neural network architecture for the encoder, decoder and latent dynamics shown previously in \cref{fig:autoencoder}. This setup was developed for a 65-species propane pyrolysis chemical kinetic model  generated automatically using \textsc{Rmg-Py}. It is emphasized that the number of network parameters is relatively low because low-latency inference is the primary objective.

\begin{table}[htbp]
\centering
\begin{threeparttable}
\caption{Detailed neural network architecture for the encoder}\label{tab:detailed_network_params1}
\centering
\begin{tabular}{C{1.6cm}C{1.4cm}C{1.9cm}}
\toprule
\toprule
\textbf{No.  of layers\tnote{$\dag$}} & \textbf{No. Neurons} & \textbf{Activations}  \\
\midrule
2 & $21 \to 3$ & Linear \\
\bottomrule
\bottomrule
\end{tabular}
\begin{tablenotes}
\footnotesize
\item[$\dag$]this includes the input layer (=21$\times$ after cull) and bottleneck layer (=3$\times$)
\end{tablenotes}
\end{threeparttable}
\end{table}

The total number of trainable network parameters is 66 for the encoder, 263 for the decoder and 4039 for the dynamics. Currently, the dimensionality of the latent space bottleneck layer is set to 3 as a trade-off between speed and accuracy; however, this can be set to any value. The non-linearities (\ie activation functions) are exponential linear units (ELU) and Gaussian linear error units (GELU), which have been shown to speed-up training and improve accuracy \cite{hendrycks2023gaussianerrorlinearunits}. However, in the output (\ie last) layer for the NN shown in \Cref{tab:detailed_network_params3}, a \textit{linear} activation function is used to avoid restricting the dynamic range of the output. For the output of the decoder, a \textit{sigmoid} non-linearity is used to bound the mass fractions between 0 and 1 according to the definition of mass fractions.

\begin{table}[htbp]
\centering
\begin{threeparttable}
\caption{Detailed neural network architecture for the decoder}\label{tab:detailed_network_params2}
\centering
\begin{tabular}{C{1.6cm}C{1.4cm}C{2.9cm}}
\toprule
\toprule
\textbf{No.  of layers\tnote{$\dag$}} & \textbf{No. Neurons} & \textbf{Activations}  \\
\midrule
6 & $3 \to 4 \to 7 \to 10 \to 20 \to 21$ & ELU (4$\times$), linear (1$\times$)  \\
\bottomrule
\bottomrule
\end{tabular}
\begin{tablenotes}
\footnotesize
\item[$\dag$]this includes the bottleneck layer (=3$\times$) and output layer (=21$\times$ after cull) 
\end{tablenotes}
\end{threeparttable}
\end{table}

The stochastic gradient descent method \texttt{AdamW} is used for optimization \cite{loshchilov2019decoupledweightdecayregularization}, and regularization is universally added using weight decay to mitigate overfitting. The backpropagated gradients through the network are clipped to prevent instability. Typically, training takes 20.4 hours on an NVIDIA Quadro RTX 6000 GPU. The training dataset is split 80\%:10\%:10\% between training, validation and test sets respectively, and the data is randomly shuffled at each training epoch. \texttt{EarlyStopping()} to used to automatically stop training once the $R^2$ score for the validation set no longer improves over a window of 10 epochs. The weights associated with the lowest $R^2$ score in the moving window are then restored.  

\begin{table}[htbp]
\centering
\begin{threeparttable}
\caption{Detailed neural network architecture for the dynamics mapping}\label{tab:detailed_network_params3}
\centering
\begin{tabular}{C{1.6cm}C{1.0cm}C{1.9cm}}
\toprule
\toprule
\textbf{No.  of layers} & \textbf{No. Neurons} & \textbf{Activations}  \\
\midrule
7 & $3 \to 48 \to 32 \to 28 \to 24 \to 20 \to 21$ & GELU (5$\times$), sigmoid (1$\times$)  \\
\bottomrule
\bottomrule
\end{tabular}
\end{threeparttable}
\end{table}


\section*{Conflicts of Interest} 
The authors declare that they have no known competing financial
interests or personal relationships that could have appeared to influence the work reported in this paper. 


\section*{CRediT Author Contribution Statement}

\textbf{Dylan Rubini}: Writing – original draft, Visualization, Validation, Software, Methodology, Investigation, Conceptualization. \textbf{Budimir Rosic}: Writing – review \& editing, Supervision, Funding acquisition, Conceptualization.  


\section*{Funding}
Dylan Rubini was supported by funding from the Engineering and Physical Sciences Research Council IAA (Grant number EP/X525777/1).

\bibliographystyle{asmems4}

\bibliography{references}

\end{document}